\documentclass[12pt]{iopart}

\usepackage{cleveref}
\usepackage{etoolbox}

\crefname{equation}{}{}

\patchcmd{\numparts}{\addtocounter{equation}{1}}{\refstepcounter{equation}}{}{}

\usepackage{iopams}
\usepackage{amssymb}
\usepackage{amsfonts}
\usepackage{graphicx}
\usepackage[T1]{fontenc}
\usepackage{textcomp}
\usepackage{xcolor}

\begin{document}

\title[]{Influence of local fields on the dynamics of four-wave mixing signals from 2D semiconductor systems }

\author{Thilo~Hahn$^{1,2}$, Jacek Kasprzak$^3$, Pawe\l ~Machnikowski$^2$, Tilmann~Kuhn$^1$, Daniel~Wigger$^2$}
\address{$^1$Institut f\"{u}r Festk\"{o}rpertheorie, Universit\"{a}t M\"{u}nster, 48149 M\"{u}nster, Germany}
\address{$^2$Department  of  Theoretical  Physics,  Wroc\l{}aw  University  of  Science  and  Technology, 50-370  Wroc\l{}aw,  Poland}
\address{$^3$Universit\'{e} Grenoble Alpes, CNRS, Grenoble INP, Institut N\'{e}el, 38000 Grenoble, France}
\ead{t.hahn@wwu.de}
\vspace{10pt}
\begin{indented}
{\it
\item[]Accepted manuscript of New J. Phys. \textbf{23} (2021) 023036 (CC BY 4.0)
\item https://doi.org/10.1088/1367-2630/abdd6c
\item[]published: 19 February 2021
}
\end{indented}

\begin{abstract}
In recent years the physics of two-dimensional semiconductors was revived by the discovery of the class of transition metal dichalcogenides. In these systems excitons dominate the optical response in the visible range and open many perspectives for nonlinear spectroscopy. To describe the coherence and polarization dynamics of excitons after ultrafast excitation in these systems, we employ the Bloch equation model of a two-level system extended by a local field describing the exciton-exciton interaction. We calculate four-wave mixing signals and analyze the dependence of the temporal and spectral signals as a function of the delay between the exciting pulses. Exact analytical results obtained for the case of ultrafast ($\delta$-shaped) pulses are compared to numerical solutions obtained for finite pulse durations. If two pulses are used to generate the nonlinear signal, characteristic spectral line splittings are restricted to short delays. When considering a three-pulse excitation the line splittings, induced by the local field effect, persist for long delays. All of the found features are instructively explained within the Bloch vector picture and we show how the exciton occupation dynamics govern the different four-wave mixing signals.
\end{abstract}

\maketitle
\section{Introduction}
With the observation of an extraordinarily strong light emission from monolayers of transition metal dichalcogenides (TMDCs) in 2010~\cite{splendiani2010NanoLett,mak2010prl}, this material class came into the focus of semiconductor optics. Since then, the research on the fundamental physics of these materials and the potential applications of them has flourished and is still growing. The optical properties of TMDCs are strongly dominated by excitons which, due to the strong Coulomb interaction in these materials, exhibit binding energies on the order of 500~meV~\cite{chernikov2014prl}, which is orders of magnitude larger than in typical III-V or II-VI semiconductors. While initially the observed spectral lines were rather broad, it turned out that by encapsulating the monolayer in hexagonal boron nitride the inhomogeneity of the structure could be strongly reduced resulting in linewidths of the excitonic transitions approaching the homogeneous limit~\cite{cadiz2017prx,jakubczyk2019acs,hayashida2020ACSaapl}. Among other techniques, also four-wave mixing (FWM) spectroscopy has been applied to these materials \cite{hao2017natcomm,jakubczyk2019acs,boule2020prm}, which gave access to the coherence and density dynamics of the excitons after ultrafast excitation. These studies revealed FWM signals even for negative delay times~\cite{boule2020prm}, which are clear hints for contributions to the signals resulting from exciton-exciton interactions. These features revived the interest in local field models for the description of FWM signals.

Exciton dynamics in semiconductor nanostructures have been thoroughly explored over many years using tools of nonlinear spectroscopy~\cite{shah2013ultrafast,mukamel1995uop,langbein2010rnc}, like differential transmission~\cite{stievater2002prb,li2003sci}, spectral hole burning~\cite{oudar1985prl,borri2000ieee} or FWM~\cite{chemla2001nat,langbein2006ol}. In the corresponding theoretical description it is often sufficient to apply two- or few-level models~\cite{allen1987optical,voss2006prb}. For systems like quantum dots (QDs), in particular in samples with a low QD density, it is generally accepted that each excitonic few-level system can be treated individually~\cite{mermillod2016opt, groll2021MQT} and a collection of QDs is then treated by ensemble averaging~\cite{vagov2004prb}. In spatially extended systems like GaAs quantum wells, however, many body effects become relevant~\cite{wegener1990pra}. It turned out that an effective mean field treatment of the exciton-exciton interaction successfully reproduces the experiments in such samples~\cite{kim1992prl,mayer1994prb}. This model describes the influence of all other excitons on the two-level system (TLS) of a single exciton in the form of a so-called {\it local field}, which allows one to analyze the resulting FWM signals in the lowest contributing order of the exciting fields, i.e., the $\chi^{(3)}$-regime~\cite{wegener1990pra}. This local field contribution to a TLS has subsequently been derived from more fundamental theoretical approaches. Starting from a microscopic density matrix description, it can be interpreted as an interaction with the exciton-exciton scattering continuum~\cite{victor1995zfb,kwong2005prb}. In the case of a resonant excitation of the 1s exciton, such local field contributions can also be obtained from a simplification of the semiconductor Bloch equations~\cite{Wegener2002semiconductor}. Recently, also in TMDC systems simplified models derived from a microscopic theory have been found to show terms which can be interpreted as local field effects~\cite{katsch2019tdm, katsch2018pssb}.

Motivated by these theoretical works and the experiments in Ref.~\cite{boule2020prm}, but not limited to TMDC systems, we build on the original model and thoroughly study the influence of the local field effect on the spectral dynamics of different FWM signals. In practice we focus on different excitation scenarios, namely a two- and a three-pulse excitation scheme, which allows us to probe different aspects of the system dynamics~\cite{mermillod2016opt}. We show that an analytic treatment in the limit of ultrafast laser pulses is possible even without restrictions to a perturbative treatment, which allows us in particular to study the dependence of the FWM signals on the intensities of the exciting pulses. An instructive explanation of the involved signal dynamics based on the Bloch vector description clarifies their physical interpretation. In turn, the numerical simulation of FWM signals generated by laser pulses with realistic durations in the sub-picosecond range allows us to make predictions for actual experiments and to study the effect of temporally overlapping pulses.

The paper is organized as follows: After the introduction to the model in Sec.~\ref{sec:model} an analytic solution of the equations of motion for ultrafast optical excitations is given in Sec.~\ref{sec:solution}. Two-pulse FWM signals are discussed in Sec.~\ref{sec:2pulse} and three-pulse signals in Sec.~\ref{sec:3pulse}, where first the $\delta$-pulse limit and then non-vanishing pulse durations are analyzed. Finally, in Sec.~\ref{sec:conclus} we draw some conclusions.

\section{Model}\label{sec:model}
The optical driving of a collection of two-level systems, each consisting of the ground state $\vert g \rangle$ and the excited state $\vert x \rangle$ separated by an energy $\hbar\omega_0$, is in general described by the optical Bloch equations~\cite{allen1987optical},
\numparts\label{eq:lfa0}
\begin{eqnarray}
	& \frac{\partial p}{\partial t} = i(1-2n)  \Omega(t)  - \beta p\,, \\
	&\frac{\partial n}{\partial t} = 2{\rm Im}[ p\Omega^*(t) ] - \Gamma n\,.
\end{eqnarray}\endnumparts
The central quantities are the polarization $p = \langle \vert g \rangle \langle x \vert \rangle$ and the occupation $n = \langle \vert x \rangle \langle x \vert\rangle$. Transitions between the two states are induced by the classical light field $\boldsymbol E(t)$, expressed here in terms of the instantaneous Rabi frequency $\Omega(t)=\boldsymbol{M}\cdot\boldsymbol{E}(t)/\hbar$. $\boldsymbol M$ is the dipole matrix element and $\boldsymbol E(t)$ is taken to be in resonance with the transition energy $\hbar\omega_0$. Therefore, both $\boldsymbol{E}(t)$ and $\Omega(t)$ are described only by their envelopes. Note that we have applied the standard rotating wave approximation (RWA) and the equations are given in the frame rotating with the transition energy $\hbar\omega_0$.The RWA is well justified for a  resonant or close-to-resonant excitation with pulse durations in the hundred femtosecond range and moderate pulse powers~\cite{kumar2011pra}. In the context of TMDC monolayers, the restriction to a two-level model is expected to be applicable, when the dynamics are restricted to excitons in a single valley, which is the case in a resonant co-circular excitation scheme. For large exciton densities the scattering into other states might become relevant~\cite{katsch2018pssb}. Nevertheless, we will study the behavior of the TLS also in the regime of larger pulse areas beyond the third order regime, keeping in mind that in actual experiments effects from other excitons might contribute to the signals. In this way, when comparing to measured signals, our results may act as a reference to estimate at which excitation powers the TLS model loses its validity. We have included a phenomenological dephasing rate $\beta$ and a decay rate of the excited state $\Gamma$. 

In the local field model the optical field is supplemented by a contribution due to the field generated by the polarization of the TLSs themselves~\cite{wegener1990pra}. The full optical field then reads $\Omega(t) = \Omega_{\rm ext}(t) + Vp$, with the external optical laser field $ \Omega_{\rm  ext}(t)$ and the coupling parameter $V$ that results from the Coulomb interaction among the excitons and that determines the strength of the self-interaction~\cite{Wegener2002semiconductor,victor1995zfb}. This substitution leads to the Bloch equations with a local field contribution,
\numparts\label{eq:lfa}
\begin{eqnarray}
	\frac{\partial p}{\partial t} &=& i(1-2n)[ \Omega_{\rm  ext}(t) + Vp ] - \beta p \,, \label{eq:lfa_p} \\
	\frac{\partial n}{\partial t} &=& 2{\rm  Im}[ p\Omega_{\rm  ext}^*(t)] - \Gamma n\,. \label{eq:lfa_n}
\end{eqnarray}
\endnumparts
Note that the local field is characterized by a real value of $V$~\cite{wegener1990pra}, therefore it does not appear in Eq.~\eref{eq:lfa_n}. Considering an imaginary part for $V$ in Eq.~\eref{eq:lfa_p} also describes an excitation induced dephasing (EID) effect~\cite{wang1994pra}, as can be derived from a microscopic theory including exciton-exciton scattering~\cite{katsch2020prl}. Note, however, that Eq.~\eref{eq:lfa_n} remains unchanged because exciton-exciton scattering does not lead to generation or recombination of excitons and therefore does not change the exciton occupation. To include both features, in the following we will treat $V$ as a complex constant. For simplicity we will continue to refer to $Vp$ as a local field effect. When discussing the analytical results we will explicitly address the influence of real and imaginary part of $V$. 

The local field obviously leads to an additional nonlinearity $\sim np$ in the optical equations. By reformulating Eq.~\eref{eq:lfa_p}, this term can be interpreted as an effective occupation-dependent shift of the transition energy from $\hbar\omega_0$ to $\hbar\omega_{\rm eff} = \hbar\omega_0 + \hbar\omega_{\rm  loc}(n)$  and an effective occupation-dependent dephasing $\beta_{\rm  eff} = \beta+ \beta_{\rm  loc}(n)$ according to
\begin{eqnarray}\label{eq:om_loc}
	\frac{\partial p}{\partial t} &=& i(1-2n)\Omega_{\rm  ext}(t) - \left[\beta + \beta_{\rm loc}(n)\right] p - i\omega_{\rm  loc}(n) p\,,\\
\end{eqnarray}
with
\begin{eqnarray}\label{eq:ombe_loc}
	\omega_{\rm  loc}(n) &=&  - (1-2n){\rm Re}(V)\,,\nonumber\\
	\beta_{\rm loc}(n) &=& (1-2n){\rm Im}(V)\,.
\end{eqnarray}
When the system is in its ground state with $n=p=0$, the effective transition frequency is reduced to $\omega_{\rm  eff}=\omega_0-{\rm Re}(V)$. It increases linearly with growing occupation, reaching $\omega_{\rm  eff}=\omega_0+{\rm Re}(V)$ in the excited state $n=1$, $p=0$. Consequently, for $n=1/2$ the frequency remains unchanged $\omega_{\rm  eff}=\omega_0$. To obtain EID, i.e., a dephasing which increases with increasing density ${\rm Im}(V)<0$ is required. This, in turn, requires $\beta > -{\rm Im}(V)$ to still have a dephasing in the linear regime.

\section{Solution for ultrashort pulses}\label{sec:solution}
Despite the nonlinearity, Eqs.~\eref{eq:lfa} can be solved analytically in the limit of ultrashort pulses, i.e., for $\delta$-pulses. Note that while mathematically we use $\delta$-pulses, physically the ultrashort pulse limit is reached if the pulse duration is much shorter than the characteristic timescale of the system's dynamics. This is still well within the validity of the RWA. As will be seen later, the obtained results will be helpful for understanding the specific behavior of the considered FWM signals. To reach the $\delta$-pulse limit, we assume a rectangularly shaped, resonant pulse of duration $\Delta t$ centered around the time $t_0$ with total pulse area $\theta$ and phase $\phi$ as
\begin{eqnarray}
	\Omega_{\rm  ext}(t) = 
	\left\{ \begin{array}{rl}
		    \hfill \displaystyle\frac{\theta e^{i\phi}}{2\Delta t}, &t_0 - \Delta t/2 \leq t \leq t_0 + \Delta t/2\,,\\
		    \hfill 0, & {\rm  otherwise\,,}
		\end{array}\right.
\end{eqnarray}
and at the end perform the limit $\Delta t \to 0$. To solve the differential equations, we use a parametrization of the instantaneous pulse area~\cite{vagov2002prb},
\begin{eqnarray}
A = \frac{\theta}{\Delta t} (t-t_0 + \Delta t/2)\,,
\end{eqnarray}
instead of the time~$t$.
Clearly, $A$ is restricted to the interval $\lbrack 0, \theta \rbrack$ corresponding to the beginning of the pulse and its end, where the total pulse area is reached. Utilizing the pulse area instead of the time avoids divergences in Eqs.~(\ref{eq:lfa}) connected with an ultrashort optical pulse. The transformed equations during the pulse read
\numparts
\begin{eqnarray}
	&\frac{\partial p}{\partial A} = i\frac{\Delta t}{\theta} (1-2n) \left(\frac{\theta e^{i\phi}}{ 2\Delta t} + Vp\right) - \frac{\Delta t}{\theta}\beta p,\\
	&\frac{\partial n}{\partial A} = {\rm  Im}(pe^{-i\phi}) - \frac{\Delta t}{\theta}\Gamma n\,.
\end{eqnarray}\label{eq:dA}\endnumparts
In the limiting case $\Delta t \to 0$ all contributions from the local field, dephasing and decay of the TLS vanish and Eqs.~\eref{eq:dA} have the same solution as the ordinary optical Bloch equations given by~\cite{vagov2002prb},
\numparts\label{eq:trafo}\begin{eqnarray}
	n^+ &=& n^- + \sin^2\left(\frac \theta2\right) (1-2 n^-) +\sin(\theta) {\rm  Im}(p^- e^{-i\phi})\,,\\
	p^+ &=& \cos^2\left(\frac\theta2\right) p^- + \frac i2 (1-2n^-) \sin(\theta) e^{i\phi} + \sin^2\left(\frac\theta2\right) p^{-*} e^{2i\phi}\,,
\end{eqnarray}\endnumparts
where $p^{+}, n^{+}$ ($p^{-}, n^{-}$) denote the polarization and occupation directly after (before) the pulse, respectively.\\

In the absence of an external field, $\Omega_{\rm  ext} = 0$, the dynamics of the occupation are only subject to the decay. For an initial occupation $n_0$ at $t=0$ it simply drops exponentially
\numparts\label{eq:prop_n}\begin{eqnarray}
	n(t) &=& n_0e^{-\Gamma t}\,.
\end{eqnarray}
This time dependence contributes to the polarization as a time-dependent phase shift introduced by the local field coupling which therefore reads
\begin{eqnarray}
	p(t) &=& p_0 \exp\left[ i\frac{2V}{\Gamma} n_0 (e^{-\Gamma t}-1)\right] \exp(iVt - \beta t)\,.
\end{eqnarray}
\endnumparts
In most semiconductor systems, the polarization dephases on a much shorter timescale than the occupation decays~\cite{mermillod2016opt}. In this case we can take $\Gamma \to 0$ and the polarization dynamics simplify to
\begin{eqnarray}
	p(t) = p_0 \exp\left[ iV(1-2n_0)t - \beta t\right]\,,
\end{eqnarray}
where the polarization simply rotates with the effective transition frequency $\omega_{\rm  eff}(n_0)=\omega_0-(1-2n_0){\rm Re}(V)$ and it decays with the effective dephasing rate $\beta_{\rm eff}= \beta + (1-2n){\rm Im}(V)$, as given in Eq.~\eref{eq:om_loc}. However, note that for completeness in the analytical results presented in the following this limit has not been performed and the parameter $\Gamma$ is still included.

\section{Two-pulse FWM}\label{sec:2pulse}
\begin{figure}[t]
	\includegraphics[width =0.6 \columnwidth]{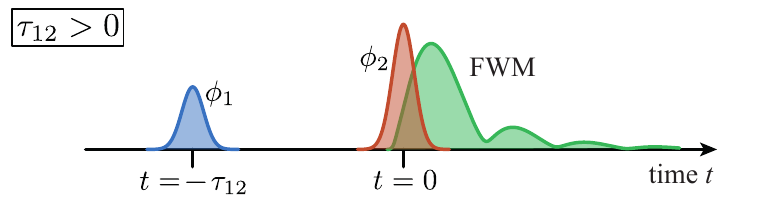}
		\caption{Pulse alignment for two-pulse FWM with positive delay $\tau_{12}>0$. The FWM signal is launched by the second pulse which also defines the time $t=0$.}\label{fig:-10}
\end{figure}

The most basic FWM experiment utilizes the two-pulse sequence depicted in Fig.~\ref{fig:-10}. Two resonant laser pulses are applied with a variable delay $\tau_{12}$ with respect to each other. The pulse areas $\theta_1$ and $\theta_2$ of the first and second pulse are kept fixed while the pulse phases $\phi_1$ and $\phi_2$ are varied during the repetition of the experiment. In the simulation the FWM signal is extracted by a phase-selection scheme from the microscopic exciton polarization as explained later. The second arriving pulse generates the FWM signal and sets the time $t=0$.
\subsection{$\delta$-pulse limit}
Based on the analytic results for a single ultrashort pulse derived in Eqs.~\eref{eq:trafo} and \eref{eq:prop_n}, we start our discussion by calculating the two-pulse FWM signal analytically in that limit. The polarization after the second pulse depends on the time $t$ after that pulse and the delay $\tau_{12}$ within the pulse pair and reads
\numparts\label{eq:poldyn}
\begin{eqnarray}
	p_2(t,\tau_{12}) &=& \Big[ a e^{i(\phi_1+ \Phi)} + b e^{i\phi_2} + c e^{i(2\phi_2-\phi_1- \Phi^{*})}  \Big] e^{ -\beta t } \nonumber\\
		&&\times\exp\Big( i\left\{\alpha + \eta e^{-{\rm Im}(\Phi)}\cos[{\rm Re}(\Phi) + \phi_1 - \phi_2] \right\}\Big) \,,\label{eq:poldyn_a}
\end{eqnarray}
with
\begin{eqnarray}
	&a(\tau_{12})  = \frac i2 \sin\left(\theta_1\right) \cos^2\left(\frac{\theta_2}2\right)\exp\left(- \beta \tau_{12} \right) \,,\label{eq:poldyn_b}\\
	&b(\tau_{12}) = \frac i2 \sin\left(\theta_2\right)\left[ 1-2\sin^2\left(\frac{\theta_1}2\right)e^{-\Gamma\tau_{12}}\right] \,,\\
	&c(\tau_{12}) = -\frac i2 \sin\left(\theta_1\right) \sin^2\left(\frac{\theta_2}2\right) \exp\left(-\beta \tau_{12} \right) \,,\\
	&\alpha(t,\tau_{12}) = Vt - 2 \frac V\Gamma \left[ \sin^2\left(\frac{\theta_2}2 \right) + \cos\left(\theta_2\right) \sin^2 
	\left(\frac{\theta_1}2 \right) e^{-\Gamma \tau_{12}} \right]\nonumber\\
	&\qquad\times\left(1-e^{-\Gamma t} \right),\\
	&\Phi(\tau_{12}) = - 2\frac V\Gamma \sin^2\left(\frac{\theta_1}2\right)\left(1 - e^{-\Gamma \tau_{12} } \right)+V \tau_{12} \,,\label{eq:poldyn_f}\\
	&\eta(t,\tau_{12}) = - \frac V\Gamma \sin\left(\theta_1\right)\sin\left(\theta_2\right) e^{-\beta \tau_{12} } \left(1 - e^{-\Gamma t}\right) \,. \label{eq:poldyn_g}
\end{eqnarray}
\endnumparts
Note, that the polarization is not oscillating with the transition frequency because we are working in the rotating frame of the laser field.
Directly after the second pulse at $t=0$, the polarization consists of three parts depending on the phases $\phi_1, \phi_2$ and $2\phi_2-\phi_1$. In the field-free propagation after the second pulse, the polarization receives an additional contribution from the occupation via the local field coupling. The corresponding term is the last exponential function in Eq.~\eref{eq:poldyn_a}, which also contributes to the phase dependence via the difference $\phi_2-\phi_1$ and its complex conjugate phase. This will be discussed in detail below. From Eqs.~\eref{eq:poldyn} we find that in the limit $\Gamma \to 0$ an imaginary part of the local field parameter only gives rise to an additional damping, depending on the pulse areas and the delay time. Thus, we expect that EID manly gives rise to a faster decay of the signals and a broadening of the corresponding spectra. Since here we are mainly interested in characteristic features in the signals and the spectra in the following we will neglect an imaginary part and assume that $V$ is a real parameter. In the \ref{sec:A2} we will show results of calculations including EID, which will confirm that this expectation is indeed fulfilled.

The FWM dynamics are obtained by filtering the polarization in Eq.~(\ref{eq:poldyn_a}) with respect to the inverse FWM phase $-\phi_{\rm  FWM}^{(2)} = -(2\phi_2-\phi_1$) and we retrieve the two-pulse FWM polarization
\begin{eqnarray}\label{eq:filter}
	p_{\rm  FWM}^{(2)}(t,\tau_{12}) = \int\limits_0^{2\pi}  p_2(t,\tau_{12}) e^{-i(2\phi_2-\phi_1)}\frac{{\rm d}\phi_1 {\rm d}\phi_2}{(2\pi)^2}\,.
\end{eqnarray}
Hence, the FWM dynamics consist only of that part of the polarization which carries the FWM phase as all other phase-dependencies vanish. This procedure models the typical heterodyne detection in FWM experiments~\cite{langbein2010rnc} Without local field interaction there is no additional phase in the free propagation. Then the FWM polarization consists only of the term $\sim c$ in Eq.~\eref{eq:poldyn_a} which reads
\begin{eqnarray}\label{eq:TLS_deph}
	p_{\rm  FWM}^{(2)}(t,\tau_{12}) = ce^{-\beta t} \sim e^{-\beta (t+\tau_{12})} \,.
\end{eqnarray} 
With local field interaction, the resulting FWM dynamics are
\begin{eqnarray}\label{eq:2pfwm}
	&p_{\rm  FWM}^{(2)}(t, \tau_{12}) = e^{i(\alpha-\Phi)-\beta t} \left[ c J_0(\eta) + ib J_1(\eta) - aJ_2(\eta)\right]  \,.
\end{eqnarray}

\begin{figure}[t]
	\includegraphics[width = 0.6\columnwidth]{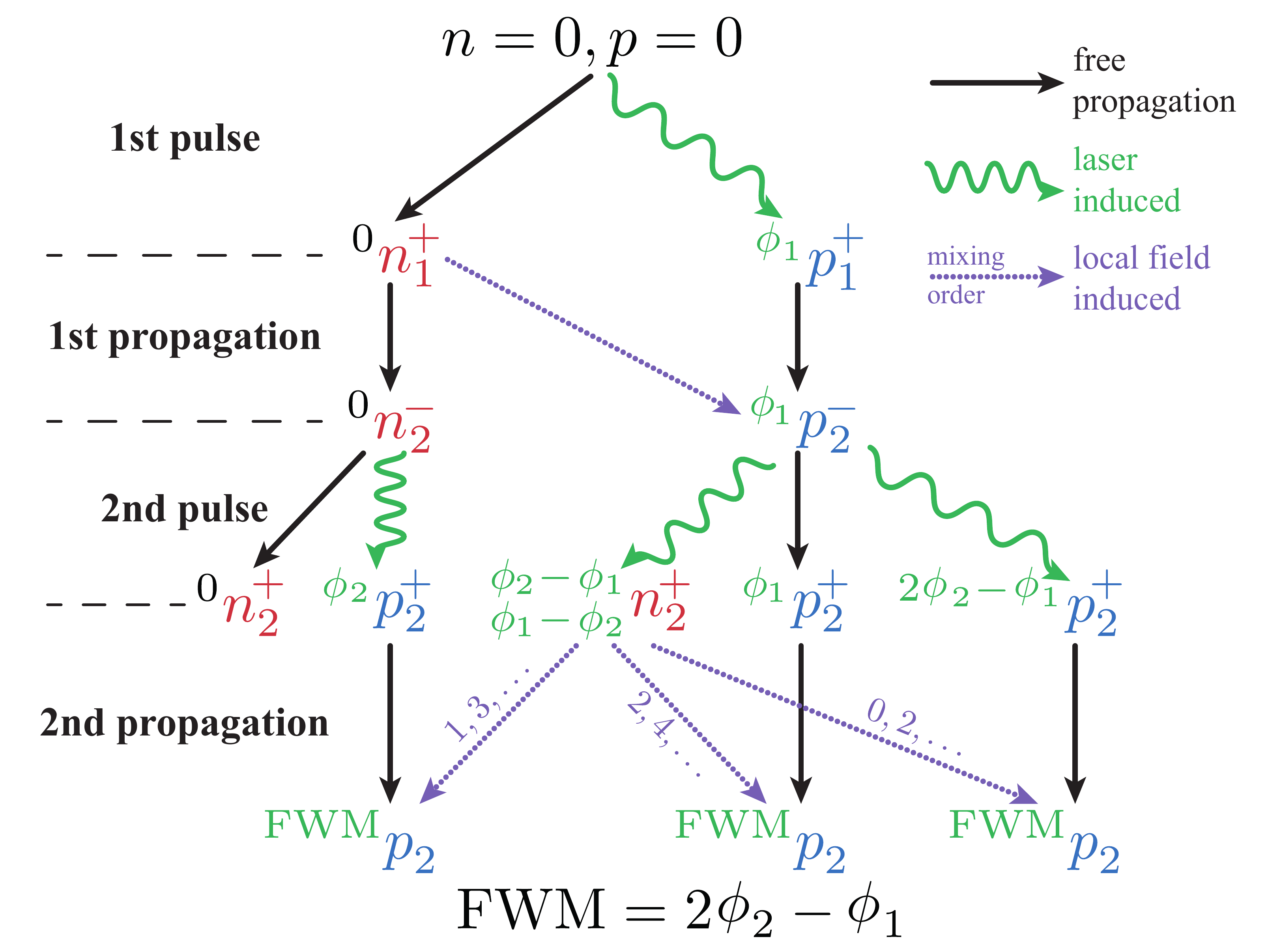}
		\caption{Flow chart of the pulse sequence and phase selection leading to the FWM signal. Black solid line: Free propagation without pulse interaction. Green wave: laser induced propagation. Violet dotted line: Local field induced mixing of the occupation into the polarizations. The numbers mark the possible mixing orders.}\label{fig:00}
\end{figure}

We find that each term of the polarization contributes to the FWM dynamics with a Bessel function of a different order. The full expression, after inserting Eqs.~\eref{eq:poldyn_b}-\eref{eq:poldyn_g} into Eq.~\eref{eq:2pfwm}, becomes quite involved, but we will now have a more detailed look on the creation of the FWM signal to understand the origins of its different contributions. A flow chart of the pulse sequence and phase selection resulting in the FWM signal is depicted in Fig.~\ref{fig:00}. Initially the system is in the ground state with $n=p=0$. The first pulse creates an occupation $^0n_1^+$ and a polarization with the phase of the first pulse $^{\phi_1}p_1^+$, marked as green wave for a laser pulse induced step. For clarity, we give the phase dependence of each quantity on the left side of the symbol and the number of the pulse on the right side, where $+$ is directly after and $-$ directly before the respective pulse. During the field-free propagation before the second pulse two things happen. On the one hand both quantities simply propagate in time indicated by black arrows. On the other hand the occupation contributes to the phase through the local field coupling  [$\Phi$ in Eq.~\eref{eq:poldyn_f}], indicated by a violet dotted arrow. The second pulse creates an occupation consisting of a part independent of the pulse phases $^0n_2^+$ and a part depending on the pulses' phase difference $^{\phi_2-\phi_1}_{\phi_1-\phi_2}n_2^+$. Because the occupation is a real quantity, this quantity has to depend on the phase difference and its conjugate, therefore two terms appear on the left side of the symbol. As mentioned above, the polarization created by the last pulse consists of parts depending on the phases of the exciting pulses $^{\phi_1}p_2^+, ^{\phi_2}p_2^+$ and the FWM phase $^{2\phi_2-\phi_1}p_2^+$. After the second pulse, $^{\phi_2-\phi_1}_{\phi_1-\phi_2}n_2^+$ contributes to the polarization via the local field resulting in additional wave-mixing possibilities (violet dotted arrows). We illustrate this by one example:

The simplest wave-mixing process between the occupation $^{\phi_2-\phi_1}_{\phi_1-\phi_2}n_2^+$ and the polarization $^{\phi_2}p_2^+$ that results in the FWM phase is
\begin{eqnarray}
	\phi_{\rm  FWM} = \underbrace{\phi_2}_{p} + \underbrace{(\phi_2-\phi_1)}_{n} \,,
\end{eqnarray}
where the occupation enters once. It is possible to generalize this to larger numbers of admixtures of $^{\phi_2-\phi_1}_{\phi_1-\phi_2}n_2^+$ giving
\begin{eqnarray}
	\phi_{\rm  FWM} = \phi_2 + k(\phi_2-\phi_1) + (k-1)(\phi_1-\phi_2)\,,
\end{eqnarray}
with $k\in\mathbb{N}$, where the occupation is used in total $2k-1$  times. In the flow chart this order of the phase difference mixing is annotated on the violet dotted lines. Considering the limit of weak local field coupling $|V|\ll \beta$, i.e., $|\eta|\ll1$ the mixing orders enter as $\eta^{2k-1}$. This directly determines the order of the corresponding Bessel function, which is one in this case. Applying this procedure also to the other polarizations we find the Bessel functions of 0th, 1st, and 2nd order.

\begin{figure}[t]
\includegraphics[width = 0.6\columnwidth]{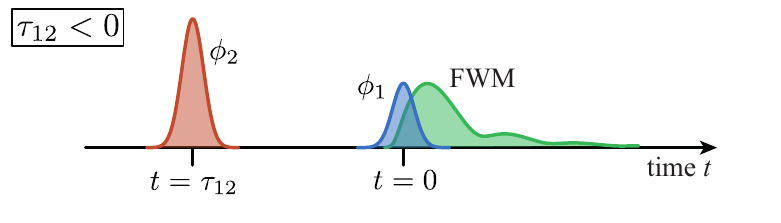}
\caption{Pulse alignment for negative delays $\tau_{12}<0$. The pulse with $\phi_1$ launches the FWM signal via the local field contribution and sets $t=0.$}\label{fig:2}
\end{figure}

In the case of a negative delay, the pulse alignment is inverted as depicted in Fig.~\ref{fig:2}. The chronologically first (second) pulse has the pulse area $\theta_2$ ($\theta_1$) and the phase $\phi_2$ ($\phi_1$). To obtain the FWM signal for negative delays, Eq.~(\ref{eq:poldyn}) can be re-used where the pulse areas, pulse phases and the sign of the delay have to be switched $\theta_1 \leftrightarrow \theta_2$, $\phi_1\leftrightarrow\phi_2$, $\tau_{12} \to \tau_{21} = -\tau_{12}$. The FWM polarization then reads
\begin{eqnarray}\label{eq:neg_delay}
	p_{\rm  FWM}^{(2)}(t, \tau_{12}<0)  =& e^{i(\alpha + 2\Phi)-\beta t}  [ ia J_1(\eta) - bJ_2(\eta) -ic J_3(\eta) ] \,.
\end{eqnarray}
In contrast to the FWM dynamics with positive delay, the orders of the Bessel functions are different as the wave-mixing orders of $^{\phi_2-\phi_1}_{\phi_1-\phi_2}n_2^+$ change. In particular, as both pulses change their ordering, in the pure TLS only the polarization $^{2\phi_1-\phi_2}p_2^+$ is created, which results in a vanishing FWM signal. Therefore, to obtain a signal with the FWM phase ${2\phi_2-\phi_1}$ at least a third order mixing via the local field is necessary. Thus, the lowest order of wave-mixing is one, which is the contribution of $^{\phi_2}p_2^+$ mixing with the phase difference $\phi_2-\phi_1$ of the occupation once.

\subsection{Simulations for non-vanishing pulse durations}
To study FWM signals with non-vanishing pulse durations $\Delta t > 0$, the Bloch equations with local field coupling in Eqs.~(\ref{eq:lfa}) and the phase filtering are evaluated numerically~\cite{haug2008quantum}. The optical field is modeled as resonant Gaussian pulses centered at $t=-\tau_{12}$ and $t = 0$,
\begin{eqnarray}
	\Omega_{\rm  ext}(t) =& \frac1{\sqrt{8\pi}\Delta t} \Bigg\{ \theta_{{}^1_2} \exp\left[{-\frac{(t+|\tau_{12}|)^2}{2\Delta t^2}+i\phi_{{}^1_2}}\right] \nonumber\\
			&\quad\qquad+ \theta_{{}^2_1} \exp\left({-\frac{t^2}{2\Delta t^2}+ i\phi_{{}^2_1}}\right)\Bigg\} \,,\\
	&{\rm  for}\quad \tau_{12} \gtrless 0\,.\nonumber
\end{eqnarray}
To account for the inverted pulse ordering, for $\tau_{12}>0$ the upper indices in $\theta_{{}^1_2}$ and $\phi_{{}^1_2}$ are used, while the lower ones refer to $\tau_{12}<0$. For the following discussion we keep as many system parameters as possible fixed to keep the analysis as instructive as possible. For the exciton dephasing we choose $\beta = 0.5$~ps$^{-1}$ which is a typical value for hBN-encapsulated TMDC monolayer~\cite{martin2020prappl}. The exciton decay time is usually much longer than the dephasing~\cite{robert2016prb} and we choose $\Gamma = 0$. For a non-vanishing value the impact of the local field effect would additionally decay on time in a non-trivial way. To avoid this complication we disregard the decay. For the pulse duration we assume $\Delta t = 0.1$~ps as a typical value for pulse trains in FWM experiments~\cite{boule2020prm}. In addition we set $\theta_2 = 2\theta_1$ which is usually applied in FWM experiments~\cite{wigger2017prb}.
As local field coupling we consider $V=6$~ps$^{-1}$ which is an order of magnitude larger than the dephasing rate $\beta$. In the literature similar proportions have been used to describe semiconductor quantum well systems~\cite{mayer1994prb, kim1992prl}. In monolayer TMDCs, the dephasing and decay rates differ between samples. Additionally, the strength of the local field is generally not known and may depend on the ambient conditions. For simplicity, it is fixed to a value such that local field effects are clearly visible. Future comparisons to experiments will help to find the correct values for the local field strength.

\begin{figure}[t]
	\includegraphics[width = 0.6\columnwidth]{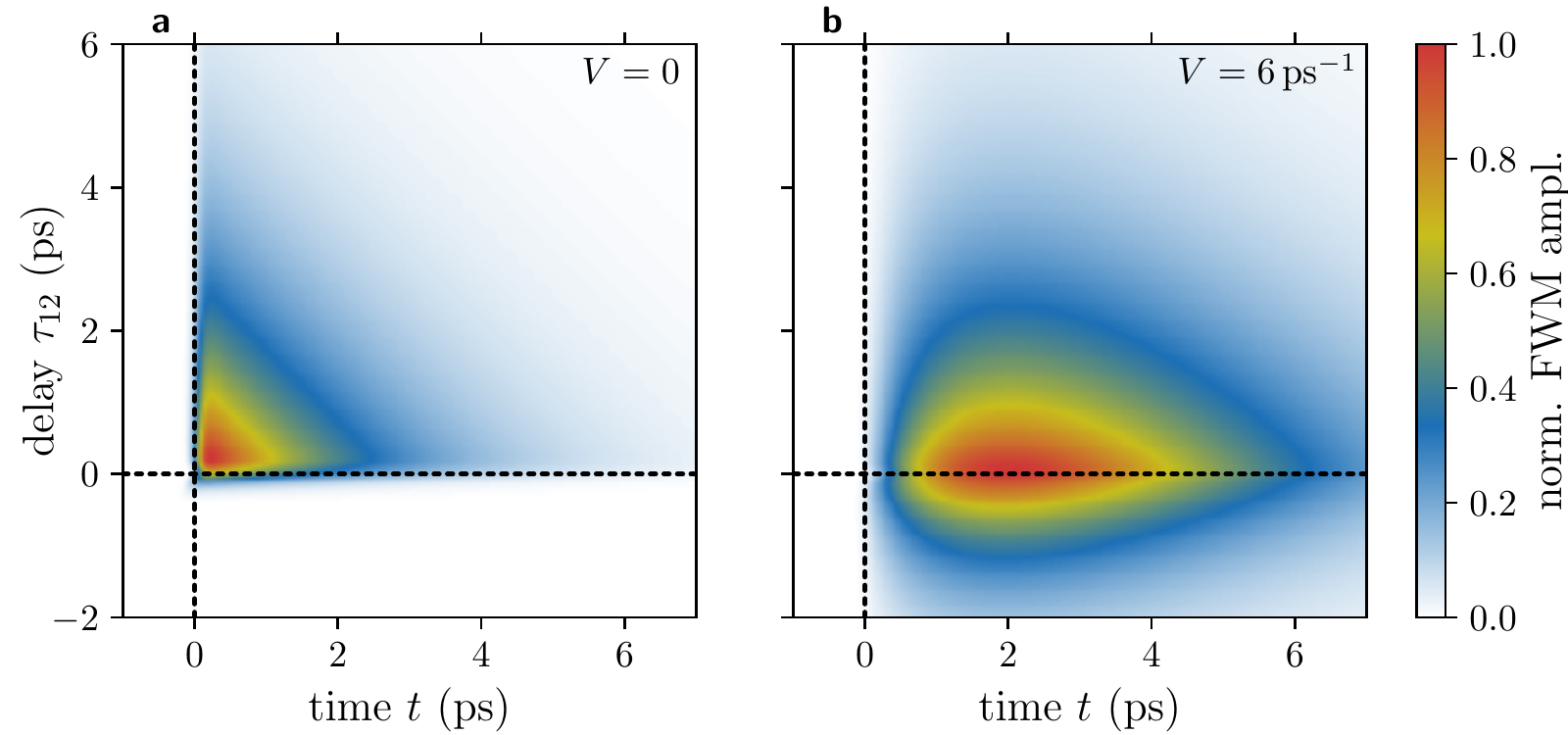}
		\caption{FWM dynamics as a function of real rime $t$ and delay $\tau_{12}$ for small pulse areas $\theta_1 = 0.05\pi$. (a) Without a local field coupling $V = 0$ and (b) with a local field coupling $V=6$~ps$^{-1}$.}\label{fig:5}
\end{figure}

We begin the discussion by analyzing the influence of the local field coupling on FWM signals in the regime of small pulse areas. Exemplarily, in Fig.~\ref{fig:5}, the FWM dynamics for $\theta_1 = 0.05\pi$ of the pure TLS with $V=0$ and with a local field coupling $V=6$~ps$^{-1}$ are shown in (a) and (b), respectively.

In the pure TLS without local field interaction, the FWM dynamics in Fig.~\ref{fig:5}(a) directly represent the polarization dynamics which are given by exponential decays in real time $t$ and delay time $\tau_{12}$ with the dephasing rate $\beta$ [Eq.~\eref{eq:TLS_deph}]. In the direction of negative delays the dynamics drop sharply when going to $\tau_{12} < 0$ as there is no FWM polarization created when the pulse with phase $\phi_2$ arrives before the pulse with phase $\phi_1$~\cite{mermillod2016opt}. The corresponding FWM spectrum, defined as
\begin{eqnarray}
	S_{\rm  FWM}(\omega,\tau_{12}) = \left|\, \int\limits_{-\infty}^\infty p_{\rm  FWM}^{(2)} (t,\tau_{12}) e^{i(\omega-\omega_0) t} {\rm d}t\,\right|
\end{eqnarray}
is a Lorentzian line for positive delays, i.e., a single peak at the bare transition energy $\hbar\omega = \hbar\omega_0$ (not shown).

Upon including the local field interaction, the FWM dynamics in Fig.~\ref{fig:5}(b) change qualitatively. On the one hand, the signal extends to $\tau_{12} < 0$ and on the other hand, the maximum of the signal is shifted to larger times $t>0$. Both aspects can be understood with the analytic solution for $\delta$-pulses. For low excitation powers, the lowest order contributing to the FWM signal is called the $\chi^{(3)}$-regime which for $t\geq 0$ reads~\cite{wegener1990pra}
\begin{eqnarray} \label{eq:chi3_2p}
	p_{\rm  FWM}^{(2)}(t,\tau_{12}) & \approx& \kappa_2  e^{iV(t- \tau_{12} )-\beta (t+ |\tau_{12}| )} \\
		&&\quad \times \left\{ \Theta(\tau_{12}) +2i Vt \left[\Theta(\tau_{12}) + e^{(-\beta+iV) |\tau_{12}| }\Theta(-\tau_{12})\right]\right\}\,, \nonumber
\end{eqnarray}
with $\kappa_2 = -i \theta_1\theta_2^2/8$ and the Heaviside function $\Theta(t)$. For positive delays, the dynamics consist of a part decaying as $e^{-\beta(t+\tau_{12})}$ which reflects simply the FWM dynamics without local field interaction and a part that rises for short times $\sim Vt e^{-\beta (t+\vert\tau_{12}\vert)}$. This contribution results from the mixing of the polarization $^{\phi_2}p_2^+$ with the phase-dependent occupation $^{\phi_2-\phi_1}_{\phi_1-\phi_2}n_2^+$, i.e., mixing order 1 in Fig.~\ref{fig:00}. In contrast, for negative delays, only the last term contributes, as the FWM polarization is not created in the pure TLS but the polarization $^{\phi_2}p_2^+$ can still mix with the occupation. Here, the pulse with the phase $\phi_2$ arrives first, therefore both $^{\phi_2}p_2^+$ and $^{\phi_2-\phi_1}_{\phi_1-\phi_2}n_2^+$ are reduced due to dephasing before the second pulse with $\phi_1$. This results in the additional decay of the signal with $e^{-\beta\vert\tau_{12}\vert}$ in Eq.~\eref{eq:chi3_2p}. So, in total, the signal decays twice as fast for negative delays than for positive ones. The FWM spectrum also consists of a single Lorentzian peak that is shifted by $-\hbar V$ due to the energy renormalization from the local field coupling (not shown).

\begin{figure}[t]
	\includegraphics[width = 0.6\columnwidth]{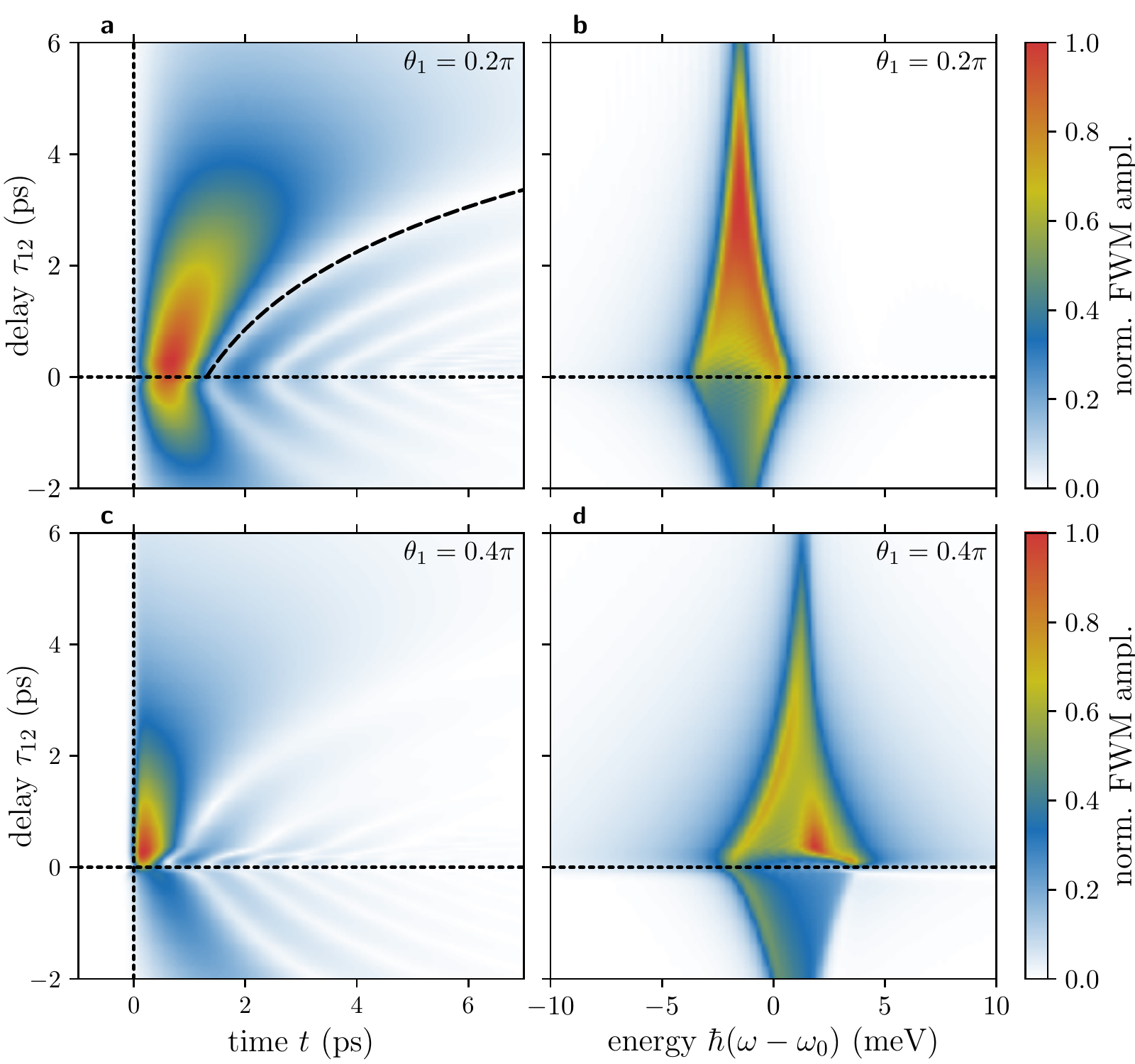}
		\caption{FWM signals for larger pulse areas. (a, b) $\theta_1 = 0.2\pi$ and (c, d) $\theta_1 = 0.4\pi$. FWM dynamics in (a) and (c) and the respective FWM spectra as a function of energy $\hbar(\omega-\omega_0)$ and delay in (b) and (d). The dashed line in (a) indicates the time dependence of the minima as predicted by Eq.~\eref{eq:minima}.}\label{fig:10}
\end{figure}

For larger pulse areas beyond the $\chi^{(3)}$-regime, the FWM dynamics change remarkably. In Fig.~\ref{fig:10}(a) the FWM dynamics  for $\theta_1 = 0.2\pi$ are shown. As a first striking difference, they exhibit oscillations, which stem from the Bessel functions known from the $\delta$-pulse solution in Eq.~\eref{eq:2pfwm}. For increasing absolute delays, the minima of this modulation shift to larger $t$. Their functional evolution $\tau_{12}^{\rm  (min)}(t)$ in the plot can be estimated from the equations. It is determined by the curve for which the argument $\eta$ of the Bessel functions [Eq. \eref{eq:poldyn_g}] is constant, i.e., its differential ${\rm d}\eta$ vanishes:
\begin{eqnarray}\label{eq:minima}
	{\rm d}\eta =\ e^{-\beta \tau_{12}} {\rm d}t - \beta t e^{-\beta \tau_{12}}{\rm d}\tau_{12} \stackrel{!}{=} 0 \nonumber\\
	\Rightarrow \ \frac{d\tau_{12}}{{\rm d}t} = \frac{1}{\beta t} \nonumber\\
	\Rightarrow \ \tau_{12}^{\rm  (min)} = \frac 1\beta \ln(t) + \tau_0 \,.
\end{eqnarray}
This curve is included as dashed line in Fig.~\ref{fig:10}(a). Therefore, the exponential loss of coherence between the pulses leads to the logarithm-shaped evolution of the minima.

\par
In Fig.~\ref{fig:10}(b), the corresponding FWM spectra are shown as a function of $\tau_{12}$. While for small absolute delays $|\tau_{12}| \ll1/\beta$, the spectrum is broad it becomes narrow for large delays, which can be explained by considering the FWM dynamics. Because the oscillations vanish for large $\tau_{12}$, the single maximum at $t>0$ leads to a single line in the spectrum. This behavior of the FWM spectra is similar for positive and negative delays and the spectral positions of large $|\tau_{12}|$ coincide. However for negative delays the signal is weaker and decays faster, as already seen in the $\chi^{(3)}$-regime in Eq.~\eref{eq:chi3_2p}. The physical meaning of the spread and position of the spectrum will be discussed below.

Also the FWM dynamics for higher pulse areas of $\theta_1 = 0.4\pi$, shown in Fig.~\ref{fig:10}(c), exhibit similarities with the previous example like the oscillatory behavior. However, one difference is that the signal decays much faster. In addition, when looking at small delays $\vert \tau_{12}\vert \approx \Delta t$, the minima deviate from the logarithmic shape. This indicates, that there is a difference between the action of two separate and two simultaneous pulses. The reason is that the pulse area and duration have a non-trivial impact on the resulting state when the system deviates from the pure TLS \cite{wigger2017prb, slepyan2004prb}. Therefore, the pulse area theorem \cite{allen1987optical} does not hold for extended pulses anymore~ \cite{slepyan2004prb} and the pulse area $\theta$ does not agree with the rotation angle of the Bloch vector. This makes the dynamics of the signal more involved. Details about this pulse area renormalization are given in \ref{sec:A1}. In addition, during the pulse overlap complex dynamics stemming from Rabi oscillations contribute to the FWM signal for large enough pulse areas~\cite{wigger2018opt}. Qualitatively also the FWM spectra, shown in Fig.~\ref{fig:10}(d), have a similar form as the ones at lower pulse area. The most striking difference is however, that the final peak energy for long delays moves from $\omega<\omega_0$ to $\omega>\omega_0$. A discussion of the differences between the simulations and the $\delta$-pulse limit is presented in \ref{sec:A2}, where the respective signals in the ultrashort pulse limit are shown. As expected, some deviations between the simulations with non-vanishing pulse duration and $\delta$-pulses appear for delays that are shorter or of the order of the pulse duration, i.e., in the regime $\tau_{12} \lesssim 200~{\rm fs}$, where the overlap has a significant influence on the generated signal. For longer delays, the $\delta$-pulse limit turns out to be a very accurate approximation.

\begin{figure}[t]
	\includegraphics[width = 0.32\columnwidth]{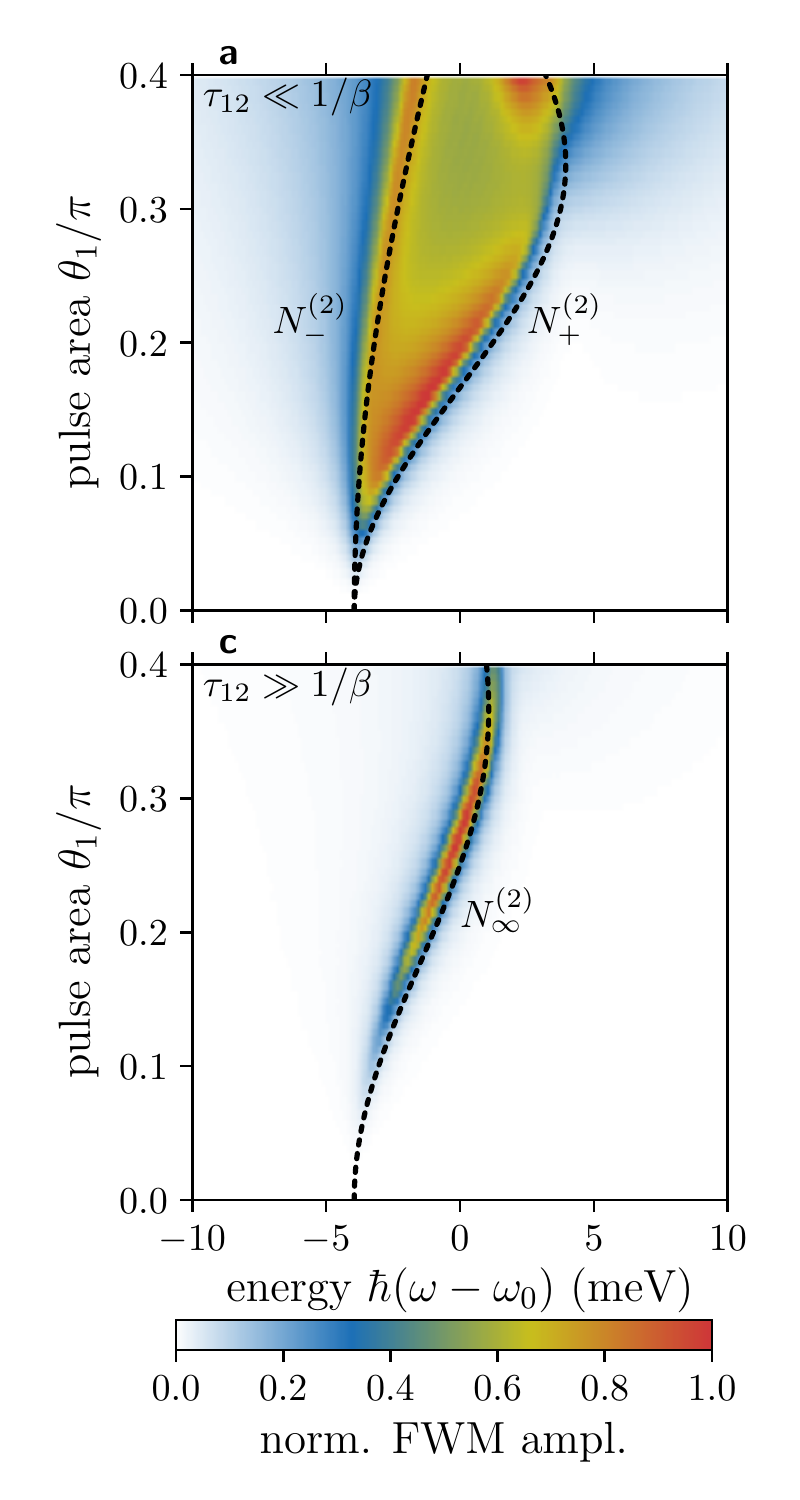}
	\includegraphics[width = 0.28\columnwidth]{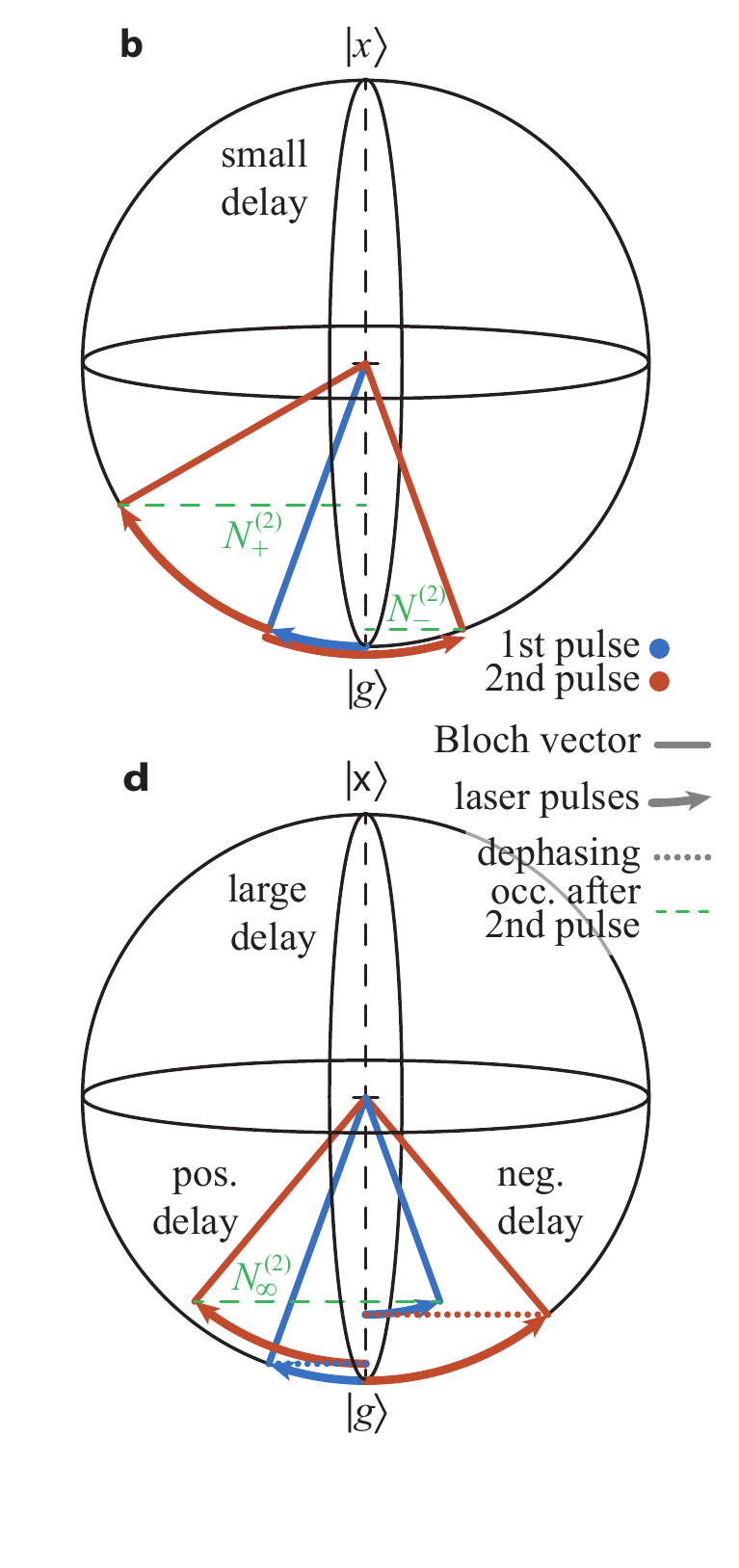}
		\caption{Limiting cases for the FWM spectra. (a) FWM spectra for small delays $\tau_{12} = 2\Delta t$ as a function of pulse area $\theta_1$. (b) Bloch vector illustration of smallest and largest energies in the spectral distribution in (a). (c) FWM spectra for long delays $\tau_{12} \gg 1/\beta$. (d) Bloch sphere illustration for (c).}\label{fig:15}
\end{figure}

In the following we analyze the spread and position of the FWM spectra at small and large delays in detail. A systematic study for small delays is shown in Fig.~\ref{fig:15}(a), where the spectra are depicted for a continuous variation of the pulse area $\theta_1$. The delay is fixed to twice the pulse duration $\tau_{12} = 2\Delta t$, such that the two pulses are mainly separated. At low pulse areas the spectrum is very narrow as we have discussed for the $\chi^{(3)}$-regime. We have already seen in the exemplary FWM spectra that the spectral width increases with growing pulse area. This spread of the spectra can be traced back to the fact, that the transition energy is modified by the occupation [see Eq.~\eref{eq:om_loc}]. So, in order to develop an intuitive picture which occupations can be reached in the FWM experiment, we consider the Bloch vector representation of the systems state in Fig.~\ref{fig:15}(b). The Bloch vector is given by its coordinates $[2{\rm Re}(p), 2{\rm Im}(p), 2n-1]$ and it points to the surface of the Bloch sphere for pure states and is shorter than unity when dephasing happens. Laser pulse excitations lead to rotations of the Bloch vector around an axis that depends on the phase of the pulse and, in general, also on the laser pulse detuning. However, here we assume a resonant excitation throughout this study.

Keeping in mind that all possible combinations of pulse phases $\phi_1$ and $\phi_2$ that match the FWM phase are realized during the repetition of the experiment, we search for the smallest and largest possible occupation after the two pulses. Following Eq.~\eref{eq:om_loc} the corresponding renormalized energies can be calculated to $\hbar\omega_{\rm  eff}$ which connects the spread of the FWM spectrum with the extent to which the system can be addressed by its coherences. We can interpret this reasoning as a kind of coherent control experiment~\cite{heberle1995prl}. In particular, such a manipulation of the occupation by varying the phases of optical pulses is performed in Ramsey fringe experiments~\cite{noordam1992pra}.

As schematically shown in the picture the largest occupation $N_+^{(2)}$ is reached when the two pulses rotate the Bloch vector into the same direction effectively adding up their pulse areas. The smallest occupation $N_-^{(2)}$ appears if the two pulses act in opposite directions. So for $\delta$-pulses the maximal (+) and the minimal ($-$) occupation are given by
\begin{eqnarray}
	N_\pm^{(2)} &=& \sin^2\left(\frac{\vert\theta_2\pm\theta_1\vert}2\right)\,.
\end{eqnarray}
The corresponding energies $\hbar\omega_{\rm  eff}(N_\pm^{(2)})$ are marked in Fig.~\ref{fig:15}(a) as dashed black lines. They describe the boundaries of the spectrum very well. But we find a slight mismatch between the dashed line and the edge of the spectrum. This deviation can be traced back to a renormalization of the pulse area due to the local field interaction, as we have mentioned before. While the pulse area is defined via its rotation angle in the pure TLS, the local field interaction changes this rotation angle in a non-trivial way (see \ref{sec:A1}).

The FWM spectra for large delays are depicted in Fig.~\ref{fig:15}(c). As we have already seen in Figs.~\ref{fig:10}(b, d), the spectrum becomes a single narrow line. Its energy depends on the pulse area and is the same for positive and negative delays. These findings can again be traced back to the behavior of the Bloch vector depicted in Fig.~\ref{fig:15}(d). After the first Rabi rotation the polarization dephases as indicated by a dotted line, i.e., the Bloch vector moves onto the Bloch sphere's $z$-axis. Afterwards, the second pulse leads to another Rabi rotation, which is reduced because the length of the Bloch vector is smaller than one. The spectral position of the FWM signal is then determined by the occupation after the second pulse $N_\infty^{(2)}$. For ultrafast pulses the final occupation reads
\begin{eqnarray}
	N_\infty^{(2)} &=& \sin^2\left(\frac{\theta_2}2\right)+\cos(\theta_2)\sin^2\left(\frac{\theta_1}2\right)\nonumber \\
				&=& \frac14 [2- \cos(\theta_2-\theta_1)-\cos(\theta_2+\theta_1)]
\end{eqnarray}
and is the same for both positive and negative delays as also shown in the schematic. This is the reason why the spectra lie at the same position for positive and negative delays. Because $N_\infty^{(2)}$ does not depend on the pulse phases anymore the spectrum consists of a single sharp line. In the context of coherent control this means that for delays much longer than the dephasing time the system cannot be addressed coherently anymore. Naturally, the loss of coherence also leads to a decay of the signal strength.

\section{Three-pulse FWM}\label{sec:3pulse}
While in the bare TLS the two-pulse FWM signal carries information on the coherence of the system, it is possible to measure the occupation dynamics in a three-pulse configuration with $\phi^{(3)}_{\rm  FWM} = \phi_3 + \phi_2 - \phi_1$ additionally~\cite{mermillod2016opt}. As for the influence of the local field the occupation plays a crucial role we expect interesting features to appear for this pulse sequence.
\subsection{$\delta$-pulse limit}
With three pulses there are two delays, $\tau_{12}$ between the first and second pulse and $\tau_{23}$ between the second and third pulse. Like in two-pulse FWM, the pulse areas $\theta_{1,2,3}$ are kept fixed while the pulse phases $\phi_{1,2,3}$ are scanned to generate the FWM signal. Usually, the delay $\tau_{12}$ is kept small, such that the pulses slightly overlap, and it stays fixed while the delay $\tau_{23}$ is varied.

As explained before in the limit of ultrashort pulses all properties of the system can be calculated analytically. Following the results in Eq.~\eref{eq:poldyn} and applying Eqs.~\eref{eq:trafo} we end up with the following phase dependencies immediately after the third pulse: The occupation $n_3^+$ carries the phases $\phi_2-\phi_1$, $\phi_3-\phi_1$, $\phi_3-2\phi_2+\phi_1$ and their complex conjugates, the polarization $p_3^+$ carries the phases $\phi_1$, $\phi_2$, $\phi_3$, $2\phi_2-\phi_1$, $2\phi_3-\phi_2$, $2\phi_3-\phi_1$, and $2\phi_3 - 2\phi_2+\phi_1$. In the following free propagation additional wave mixing between the occupation and the polarization can contribute to the FWM signal. As this large number of possible phase combinations complicates the situation massively, we will only analyze the limiting case of large delays $\tau_{23} \gg 1/\beta$ analytically. So we assume that before the interaction with the third pulse the polarization has dephased entirely and we can set $p_3^-=0$. Choosing further $\tau_{12} = 0$, the polarization after the three pulses reads
\numparts
\begin{eqnarray}
	&p_3(t,\tau_{23}) = \exp\left\lbrace i[\vartheta + \nu \cos(\phi_2-\phi_1) +\phi_3]-\beta t\right\rbrace \nonumber\\
		& \qquad\qquad\qquad \times\left[ u + 2 w \cos(\phi_2-\phi_1)\right] \,,
\end{eqnarray}
with
\begin{eqnarray}
	&\vartheta(t,\tau_{23}) = Vt - \frac{2V}\Gamma \left(1 - e^{-\Gamma t}\right) \bigg\{ \sin^2\left(\frac{\theta_3}2\right) + e^{-\Gamma\tau_{23}}\cos(\theta_3)\nonumber\\
		&\quad \times \left[ \sin^2\left(\frac{\theta_2}2\right) + \cos(\theta_2)\sin^2\left(\frac{\theta_1}2\right)\right]\bigg\},\\
	&\nu(t,\tau_{23}) =- \frac V\Gamma \sin(\theta_1)\sin(\theta_2)\cos(\theta_3)e^{-\Gamma\tau_{23}}\left(1 - e^{-\Gamma t}\right),\\
	&u(\tau_{23}) = \frac i2 \sin(\theta_3)\bigg[ 1-2 \sin^2\left(\frac{\theta_2}2\right)\nonumber\\
		&\qquad -2 \cos(\theta)\sin^2\left(\frac{\theta_1}2\right)\bigg] e^{-\Gamma \tau_{23}},\\
	&w(\tau_{23}) = -\frac i4 \sin(\theta_1)\sin(\theta_2)\sin(\theta_3) e^{-\Gamma \tau_{23}}.
\end{eqnarray}
\endnumparts
Note that these equations are valid for a complex local field parameter $V$, i.e., EID is included. However, in the following we will again concentrate on a real parameter $V$.

\begin{figure}[b]
	\includegraphics[width =0.6 \columnwidth]{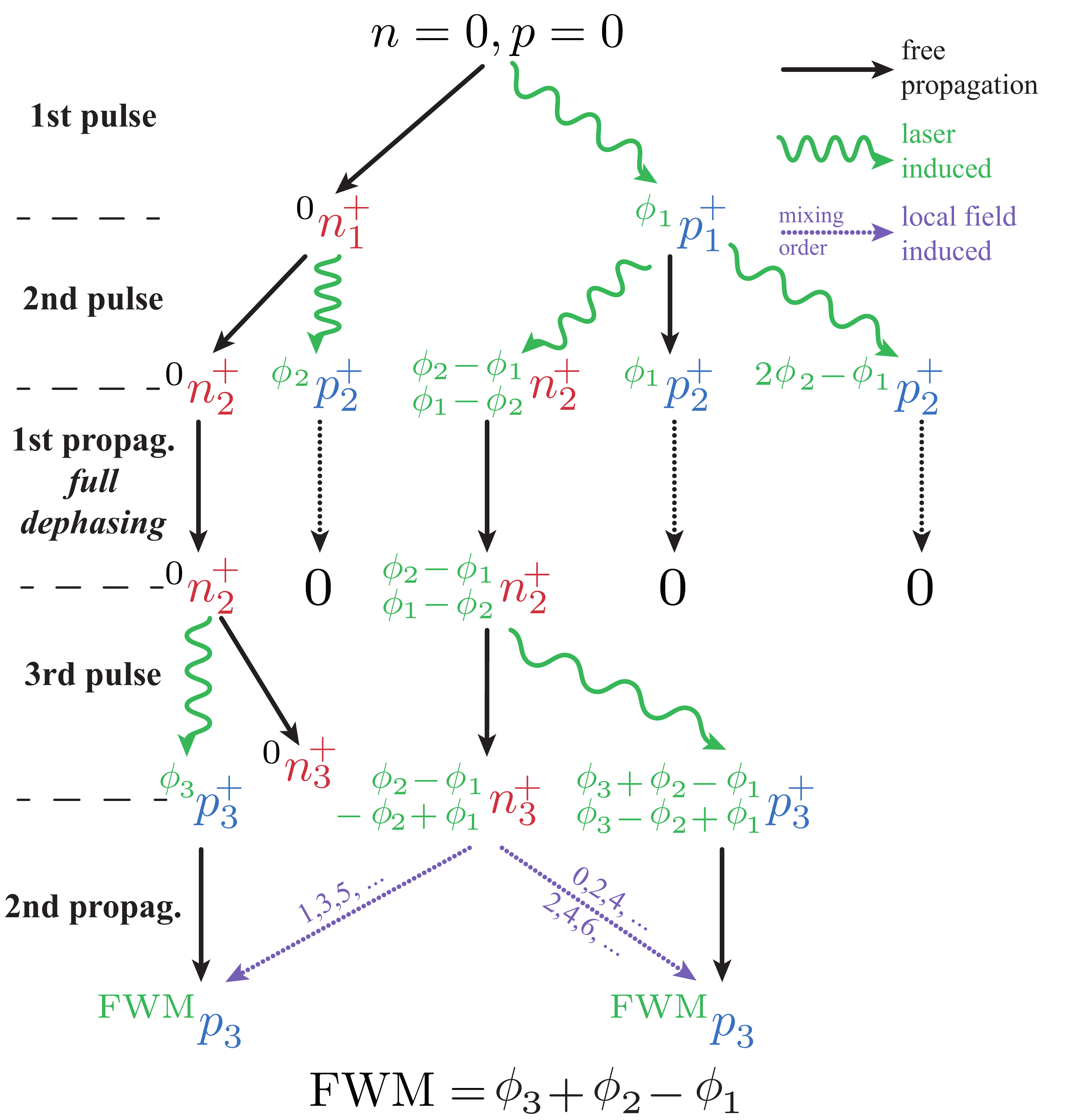}
		\caption{Flow chart of the generation of the three-pulse FWM signal. The delay $\tau_{23}$ is supposed to be long enough that all polarizations can fully dephase. The different arrows have the same meaning as in Fig.~\ref{fig:00}.}\label{fig:20}
\end{figure}

Using the FWM phase $\phi_{\rm  FWM}^{(3)} = \phi_3+\phi_2-\phi_1$, the three-pulse FWM dynamics are obtained by the phase integration
\begin{eqnarray}
	p_{\rm  FWM}^{(3)}(t,\tau_{23}) = \int\limits_0^{2\pi} p_3(t,\tau_{23}) e^{-i(\phi_3+\phi_2-\phi_1)}\frac{{\rm d}\phi_1{\rm d}\phi_2 {\rm d}\phi_3}{(2\pi)^3}\,.
\end{eqnarray}
A flow chart of the single steps leading to the FWM signal is shown in Fig.~\ref{fig:20}. Analogous to Fig.~\ref{fig:00} the first and second pulse create an occupation which consists of a part independent of the pulse phases $^0n_2^+$ and a part dependent on the pulses' phase difference $^{\phi_2-\phi_1}_{\phi_1-\phi_2}n^+_2$. These two terms remain even for long delays because we assume that the excited state decay is much slower than the dephasing. The polarization which consists of $^{\phi_1}p_2^+$, $^{\phi_2}p_2^+$ and $^{2\phi_2-\phi_1}p_2^+$ vanishes due to the long delay $\tau_{23}$ which is indicated by black dotted arrows.
As a consequence the last pulse creates polarizations exclusively from the occupations before this pulse. Therefore, the final polarization consists of a part with the phase of the third pulse $^{\phi_3}p_3^+$, indicated by the green wave from $^0n_2^+$, and a part $^{\phi_3+\phi_2-\phi_1}_{\phi_3-\phi_2+\phi_1}p_3^+$ which adapts the phase difference from the occupation $^{\phi_2-\phi_1}_{\phi_1-\phi_2}n_2^+$. There is also the possibility that the occupation does not receive an additional phase dependence and simply becomes $^{\phi_2-\phi_1}_{\phi_1-\phi_2}n_3^+$.

Without the local field coupling, only $^{\phi_3-\phi_2+\phi_1}p_3^+$ contributes to the FWM signal and we see that it carries the information of the previous occupation. Through the local field coupling, the polarization mixes with $^{\phi_2-\phi_1}_{\phi_1-\phi_2}n_3^+$ in the following field-free propagation which is marked by violet dotted arrows. This again results in Bessel functions describing the dynamics of the FWM signal. In direct analogy to the discussion of two-pulse FWM, the minimal number of mixing processes involved determines the order of the Bessel function. The possible mixing orders are again annotated on the violet arrows.

Finally, in case of long delays $\tau_{23}\gg1/\beta$, the FWM dynamics read
\begin{eqnarray}
	&p^{(3)}_{\rm  FWM} = e^{i\vartheta-\beta t}[ wJ_0(\nu) + iuJ_1(\nu) - wJ_2(\nu) ] \,.
\end{eqnarray}
The dynamics have a similar form as the analytical result from the two-pulse FWM in Eq.~\eref{eq:2pfwm}. This is because the occupation $n_2^+$ has the same phase dependence as in the two-pulse case and lowest mixing orders are also 0, 1, and 2 for $^{\phi_3+\phi_2-\phi_1}p_3^+$, $^{\phi_3}p_3^+$, and $^{\phi_3-\phi_2+\phi_1}p_3^+$, respectively.

\subsection{Simulations for non-vanishing pulse durations}
To discuss non-vanishing pulse durations $\Delta t >0$, we simulate the FWM signals numerically with Gaussian laser pulses. We set all pulse areas to the same value $\theta_1 = \theta_2 = \theta_3$ and fix the delay between the first two pulses to $\tau_{12} = 0.2$~ps. The other parameters agree with the two-pulse case, i.e., $\beta = 0.5$~ps$^{-1}$, $\Gamma = 0$, and $\Delta t = 0.1$~ps.

\begin{figure}[t]
	\includegraphics[width = 0.6\columnwidth]{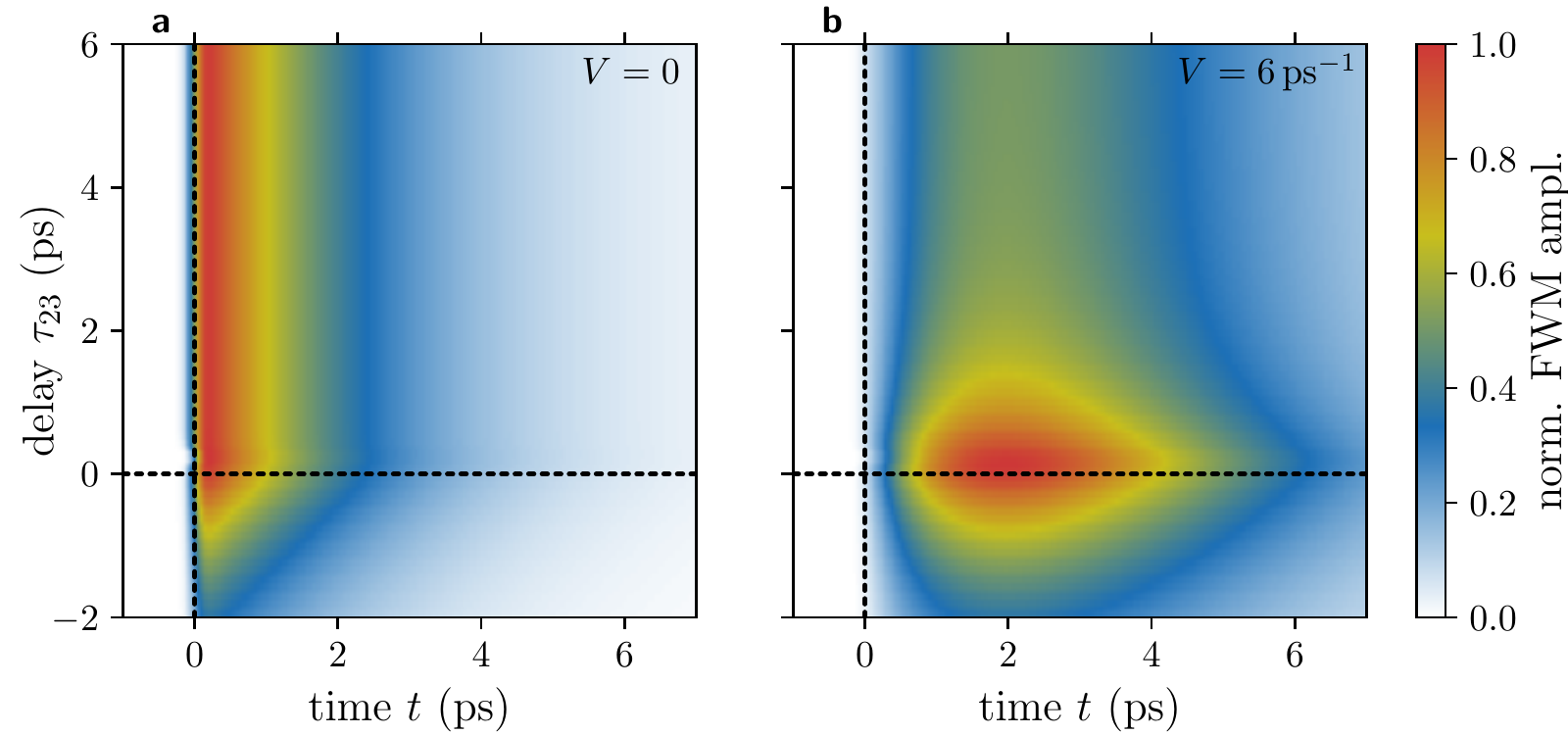}
		\caption{Three-pulse FWM dynamics for small pulse areas $\theta_1 = 0.05\pi$. (a) Without a local field coupling $V = 0$ and (b) with a local field coupling $V=6$~ps$^{-1}$.}\label{fig:25}
\end{figure}

Beginning with small pulse areas, the FWM dynamics for $\theta_1 = 0.05\pi$ without local field interaction $V=0$ and with a local field coupling of $V=6$~ps$^{-1}$ are shown in Figs.~\ref{fig:25}(a) and (b), respectively. For $V=0$ in (a) the FWM signal decays in real time $t$ as the FWM polarization dephases. The signal does not change when varying positive delays $\tau_{23}$ because we did assume a vanishing decay rate and the signal is probing the occupation dynamics. For negative delays however, the pulse with $\phi_3$ arrives first. After this pulse only the polarization is phase-dependent and the next two pulses with $\phi_1$ and $\phi_2$ create the FWM signal. Therefore the FWM signal reflects the polarization dynamics. This quantity decays $\sim e^{-\beta |\tau_{23}| }$ due to the dephasing in the field-free propagation after the single pulse.

With the local field interaction in Fig.~\ref{fig:25}(b), the signal changes qualitatively. Similar to two-pulse FWM, the maximum of the FWM dynamics is shifted to $t>0$ due to the wave-mixing in the free propagation after all three pulses. In addition the signal is strongest for short delays $\tau_{23} \ll 1/\beta$. These aspects can be explained by calculating the FWM dynamics in the $\chi^{(3)}$-regime to
\begin{eqnarray}\label{eq:3p_chi}
	p_{\rm  FWM}^{\chi^{(3)}} =& \kappa_3 e^{iVt-\beta t} 
		 \bigg\{ \Theta(\tau_{23} )\left[ 1  + i Vt \left(1 + e^{-2\beta \tau_{23} }\right) \right] \\
			&\qquad\qquad+  \Theta(-\tau_{23})e^{-\beta |\tau_{23}| + iV|\tau_{23}| }( 1  +2 iVt) \bigg\} \,,\nonumber
\end{eqnarray}
with $\kappa_3 = -i \theta_1\theta_2\theta_3/4$. Independent of the sign of the delay, the energy is renormalized. Hence, the FWM spectrum consists of a single line at $\omega = \omega_0-V$ (not shown).

For positive delays the first term in the curly brackets in Eq.~\eref{eq:3p_chi} contributes. There, the first contribution simply stems from the pure TLS without local field coupling. The second contribution is created by the interaction with the local field and consists of two parts. The first one does not depend on the delay and stems from the mixing between the polarization after the third pulse $^{\phi_3}p_3^+$ and the occupation $^{\phi_2-\phi_1}n_3^+$. The second part decays with $e^{-2\beta\tau_{23}}$ and stems from the polarization $^{\phi_2}p_3^+$ which mixes with $^{\phi_3-\phi_1}n_3^+$. As both of these quntities stem from the polarization after pulses 1 and 2, they suffer from the dephasing in the field-free propagation after the pulse pair. Therefore this contribution decays with twice the dephasing rate.

For negative delays the second term in the curly brackets in Eq.~\eref{eq:3p_chi} contributes. The pulse with phase $\phi_3$ arrives first and after the delay $\tau_{23}$, the two pulses with $\phi_1$, $\phi_2$ excite almost simultaneously. Consequently the only contributing quantity after the first excitation is $^{\phi_3}p_3^+$ and the entire signal decays with $e^{-\beta |\tau_{23}| }$. The FWM dynamics consist of a contribution from the pure TLS and a part from the local field interaction. The latter consists of the mixing between $^{\phi_3}p_3^+$ and $^{\phi_2-\phi_1}n_3^+$ and between $^{\phi_2}p_3^+$ and $^{\phi_3-\phi_1}n_3^+$ resulting in the term $\sim Vt$.

\begin{figure}[t]
	\includegraphics[width = 0.6\columnwidth]{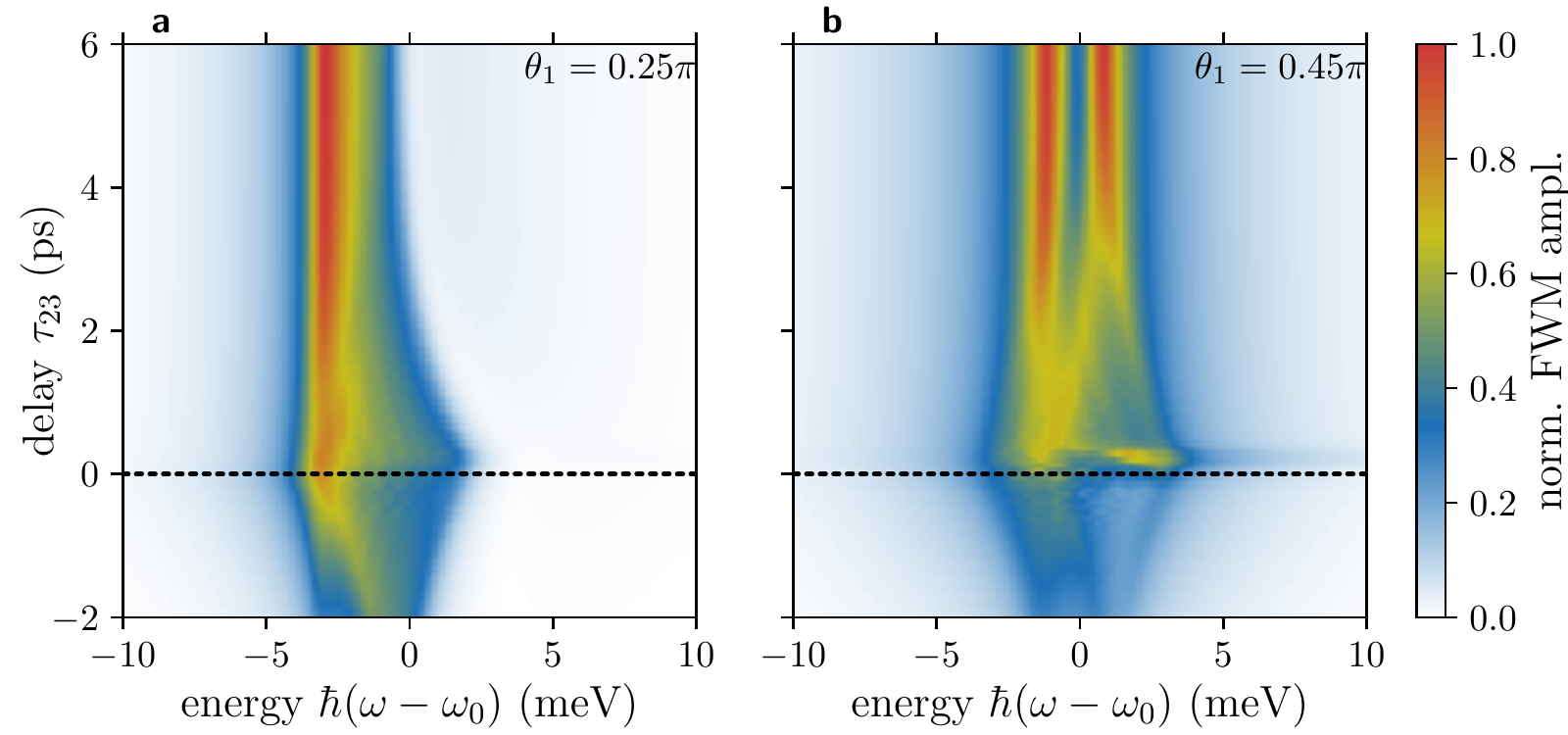}
		\caption{Three-pulse FWM spectra for larger pulse areas. (a) $\theta = 0.25\pi$ and (b) $\theta = 0.45\pi$. The local field coupling is $V=6$~ps$^{-1}$.}\label{fig:30}
\end{figure}

Going to larger pulse areas, the FWM spectra first change slightly. For an intermediate pulse area of  $\theta_1 = 0.25\pi$ in Fig.~\ref{fig:30}(a) the FWM spectrum is initially significantly broadened and becomes narrower for larger delays but it does not evolve into a single symmetric line and remains broader than the Lorentzian in the $\chi^{(3)}$-regime. A detailed discussion of the linewidth at large delays will be carried out later.

At still higher pulse areas, e.g., $\theta_1 = 0.45\pi$ depicted in Fig.~\ref{fig:30}(b), the FWM spectra change significantly. Initially, the FWM spectrum is broad  and evolves into two separate lines for large $\tau_{23}$ which is a striking difference to the previous picture. Both lines are separated by a minimum of the signal at $\omega=\omega_0$. This characteristic behavior will be explained in the context of coherent control later.

For negative delays $\tau_{23} < 0$ the details of the FWM spectra change significantly when comparing Figs.~\ref{fig:25}(a) and (b) and become quite involved. However, in all cases the intensity decays rapidly due to the dephasing as discussed for the $\chi^{(3)}$-regime.

\begin{figure}[t]
	\includegraphics[width = 0.3\columnwidth]{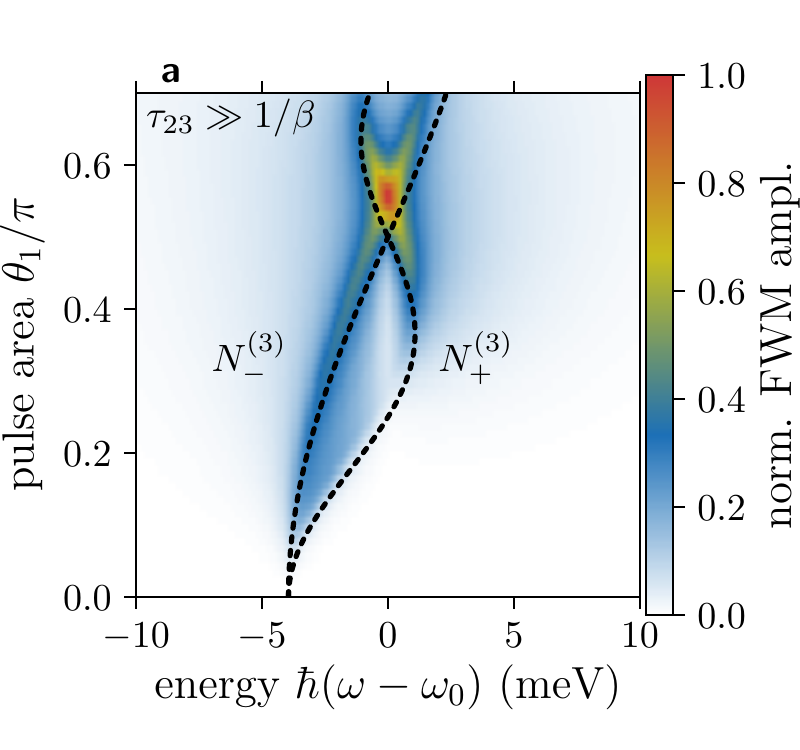}
	\includegraphics[width = 0.3\columnwidth]{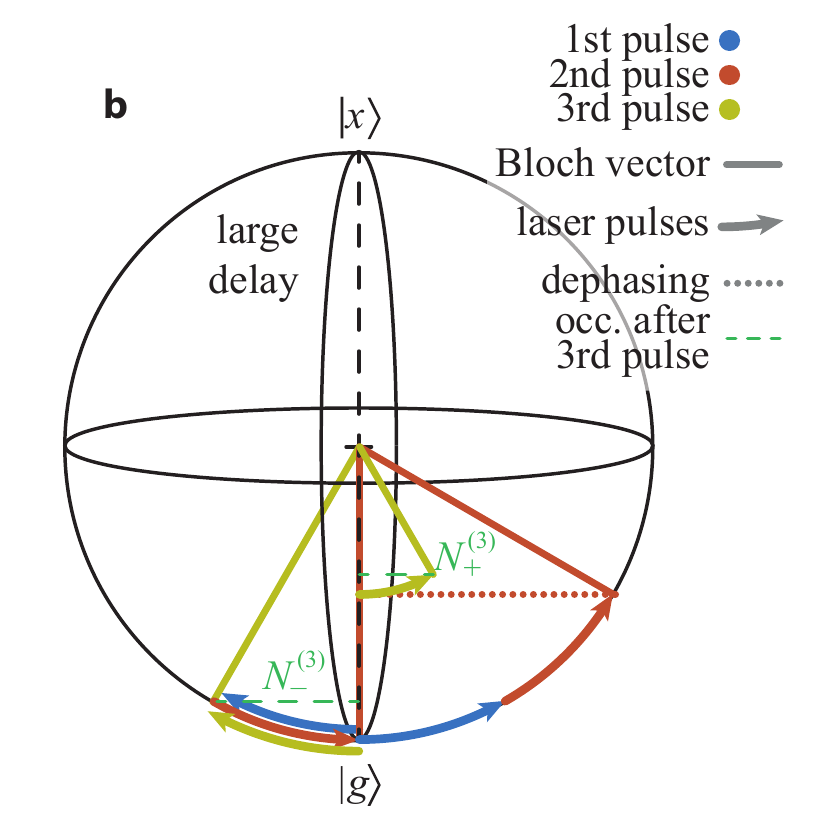}
		\caption{(a) Three-pulse FWM spectra in the limit of longe delays $\tau_{23} \gg 1/\beta$. (b) Sketch of the Bloch vector to illustrate the generation of the smallest and largest possible energy in the spectrum.}\label{fig:40}
\end{figure}

While the spectra are all broadened and quite involved for small delays $\tau_{23}\approx 1$~ps they become very clean for long delays $\tau_{23}\approx 5$~ps. The reason for this behavior is the multitude of different possible phase combinations for short delays and the vanishing influence of the polarizations for long delays as explained before. We find pulse areas that result in a single or in two separate lines. This effect is further highlighted in Fig.~\ref{fig:40}(a), where the FWM spectra for large delays $\tau_{23}\gg1/\beta$ are shown as a function of the pulse area $\theta$. For small pulse areas in the $\chi^{(3)}$-regime, the spectrum appears as a single line at $\omega = \omega_0-V$. Then the spectrum broadens for an increasing pulse area. At $\theta_1 \approx \pi/4$, the spectrum splits into two parts with a pronounced minimum at $\omega = \omega_0$ in agreement with Fig.~\ref{fig:30}(b). For $\theta_1 \approx \pi/2$, the two lines merge again into a single peak at $\omega = \omega_0$. Increasing the pulse area further, the single line splits into two lines again.

For the explanation of this behavior we consider the Bloch vector picture again. Figure~\ref{fig:40}(b) shows the schematic construction of the smallest and largest possible occupation after the three pulses and a full dephasing after pulse two. The first Rabi rotations (blue and orange) can act destructively, i.e., the respective rotations of the Bloch vector compensate each other. The state of the TLS after the second pulse then points to the south pole. The third pulse rotates from there and reaches the final occupation
\numparts
\begin{eqnarray}
	 N_-^{(3)} &=& \sin^2\left(\frac {\theta_3}2\right)\,,
\end{eqnarray}
which is the smallest possible. Note, that strictly speaking this phase combination leads to a vanishing FWM signal because both, polarization and occupation are zero after the second pulse. But at this point we are just interested in the hypothetically reached shift of the transition energy. In the other extreme case the two first pulses act constructively and the pulse areas essentially add up. Before the third pulse the state dephases entirely, which means that it is projected onto the $z$-axis. From there the third pulse rotates the Bloch vector on a smaller circle and reaches the final occupation
\begin{eqnarray}
	 N_+^{(3)} &=& \sin^2\left(\frac {\theta_3}2\right) + \cos(\theta_3)\sin^2\left(\frac{\theta_1+\theta_2}2\right)\,.
\end{eqnarray}
\endnumparts
The effective transition energies corresponding to these two limiting cases are depicted in Fig.~\ref{fig:40}(a) as dashed black lines. We find that they again very accurately follow the boundaries of the calculated spectra. The slight deviation again stems from the pulse area renormalization due to the local field coupling (see \ref{sec:A1}).

Also the reduction of the signal at $\omega = \omega_0$ (seen most clearly for pulse areas around $\theta_1 = 0.4\pi$) can be explained in this picture. We consider all possible combinations of $\phi_1$ and $\phi_2$ that make the Bloch vector point to the equator after the second pulse. The following dephasing ends in the center of the sphere, which is the balanced mixture of ground and excited state. In this point the system is called transparent because the third pulse cannot create a polarization from there. Consequently also no FWM signal can be generated.

For the special case $\theta_1 = 0.5\pi$, where the spectral peaks in Fig.~\ref{fig:40}(a) merge at $\omega = \omega_0$, we remember that after the full dephasing for long delays $\tau_{23}$ all possible Bloch vectors lie on the $z$-axis of the Bloch sphere. From there the Bloch vector is rotated by the third pulse into the equatorial plane no matter where on the $z$-axis it was. Therefore the only possible occupation is $n_3^+=1/2$ which results in a renormalized energy of $\hbar\omega_{\rm  loc}=0$ [Eq.~\eref{eq:om_loc}], i.e., a single line in the spectrum at $\omega=\omega_0$. Note, that in this scenario the equator is reached after the third and not the second pulse, so the argument is not in conflict with the previous discussion of the spectral minimum at $\omega=\omega_0$.

\section{Conclusions}\label{sec:conclus}
Motivated by recent nonlinear FWM measurements on TMDC monolayers we revived the Bloch equation model extended by a local field effect. This effect takes exciton-exciton interactions into account on a mean field level. After introducing analytic solutions for the optical driving and the field-free propagation between two laser pulses in the limit of ultrashort excitations we calculated the FWM signal after a two-pulse excitation in this limit. This result was later used to explain the numerically simulated spectral dynamics for excitations with laser pulses of non-vanishing duration. We found that the local field effect leads to oscillations in the FWM signal dynamics which translate into a spectral broadening and even a line splitting for short pulse delays. We have explained these aspects with the pulses' action on the TLS's occupation, visualized by means of the Bloch vector, which determines the FWM signal via the local field. In the case of three-pulse FWM signals, the occupation dynamics plays the central role and we found that line splittings even persist for long delays between pulses two and three. Utilizing the illustrative Bloch vector again, we showed that the variations of the FWM spectral dynamics strongly depend on the applied pulse areas.

Overall, this work develops a basic understanding of nonlinear optical signals from systems where exciton-exciton interaction can be modeled by a local field approximation. Motivated by the first experiments on TMDCs indicating local field effects, this work proposes how the fundamental physics can be explored by investigating the spectral dynamics in different FWM scenarios. Especially promising is the utilization of the local field model for situations where funneling effects lead to increased exciton densities and therefore to pronounced exciton-exciton interaction.

\ack
T.H., P.M., T.K., and D.W. acknowledge support from the Polish National Agency for Academic Exchange (NAWA) under an APM grant. T.H. thanks the German Academic Exchange Service (DAAD) for financial support (No. 57504619). D.W. thanks NAWA for financial support within the ULAM program (No. PPN/ULM/2019/1/00064).

\appendix
\section{Rabi rotations for non-vanishing pulse durations}\label{sec:A1}
\begin{figure}[h]
	\includegraphics[width = 0.6 \columnwidth]{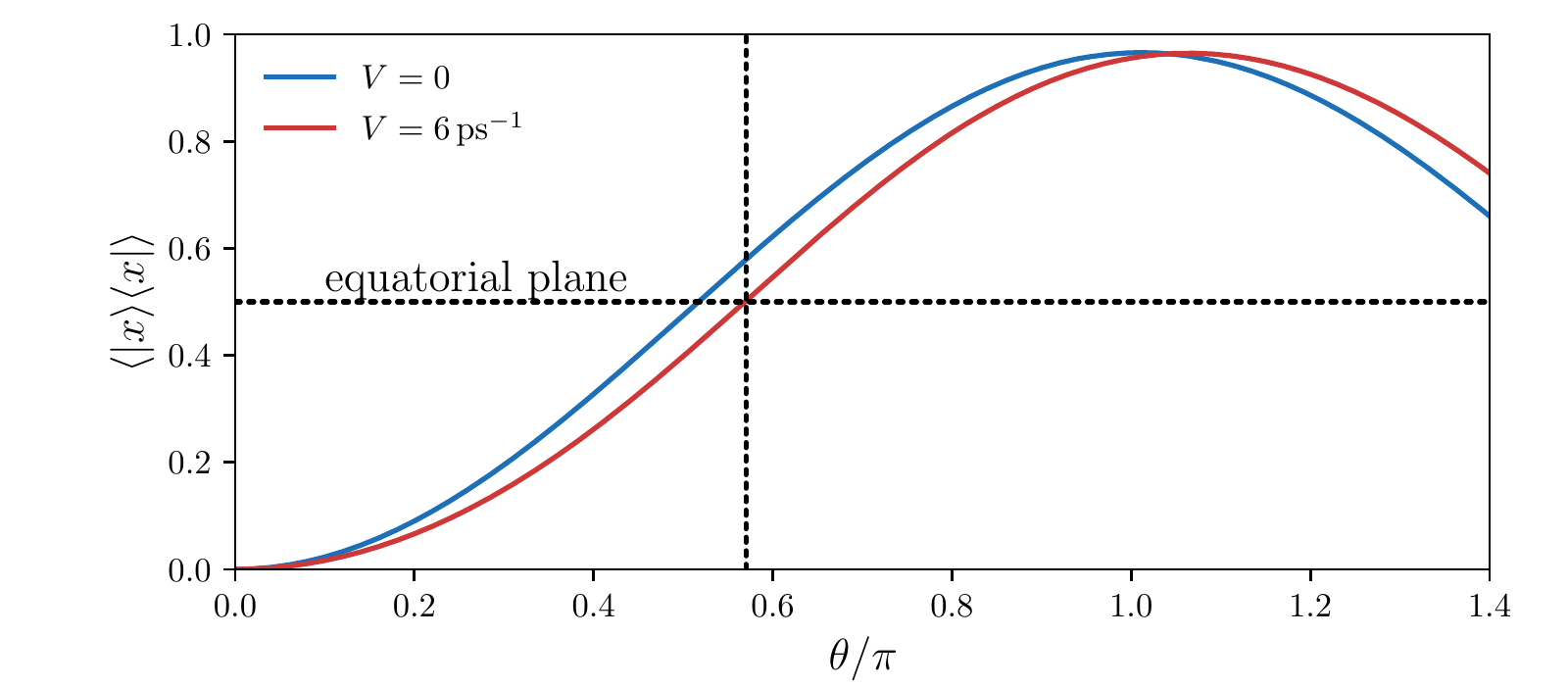}
		\caption{Occupation after a single pulse without $V=0$ (blue) and with local field interaction $V = 6$~ps$^{-1}$ (red).}\label{fig:45}
\end{figure}

In the discussion of non-vanishing pulse durations, we have noted qualitatively that the rotation angle of the Bloch vector is altered in the presence of a local field. In Fig.~\ref{fig:45} this behavior is depicted quantitatively for the parameters used in the simulations, i.e., $\beta = 0.5\,$ps$^{-1}$, $\Gamma = 0$, and $\Delta t = 0.1\,$ps. In comparison to $V = 0$ (blue line), the rotation angle for $\theta < \pi$ is smaller when a local field is taken into account (red line). This difference is largest for $\theta\approx\pi/2$. In addition, the assumed dephasing of the polarization prohibits a full occupation of the excited state in both cases. Therefore, the reduction of the effective pulse area becomes eminent for the discussion of Figs.~\ref{fig:15} and \ref{fig:30}, where the spectral features for extended pulses lie slightly above the $\delta$-pulse results. Both would agree when taking the effective pulse area into account.

\section{FWM signals for ultrashort pulses}\label{sec:A2}
\begin{figure}[t]
	\includegraphics[width = 0.6 \columnwidth]{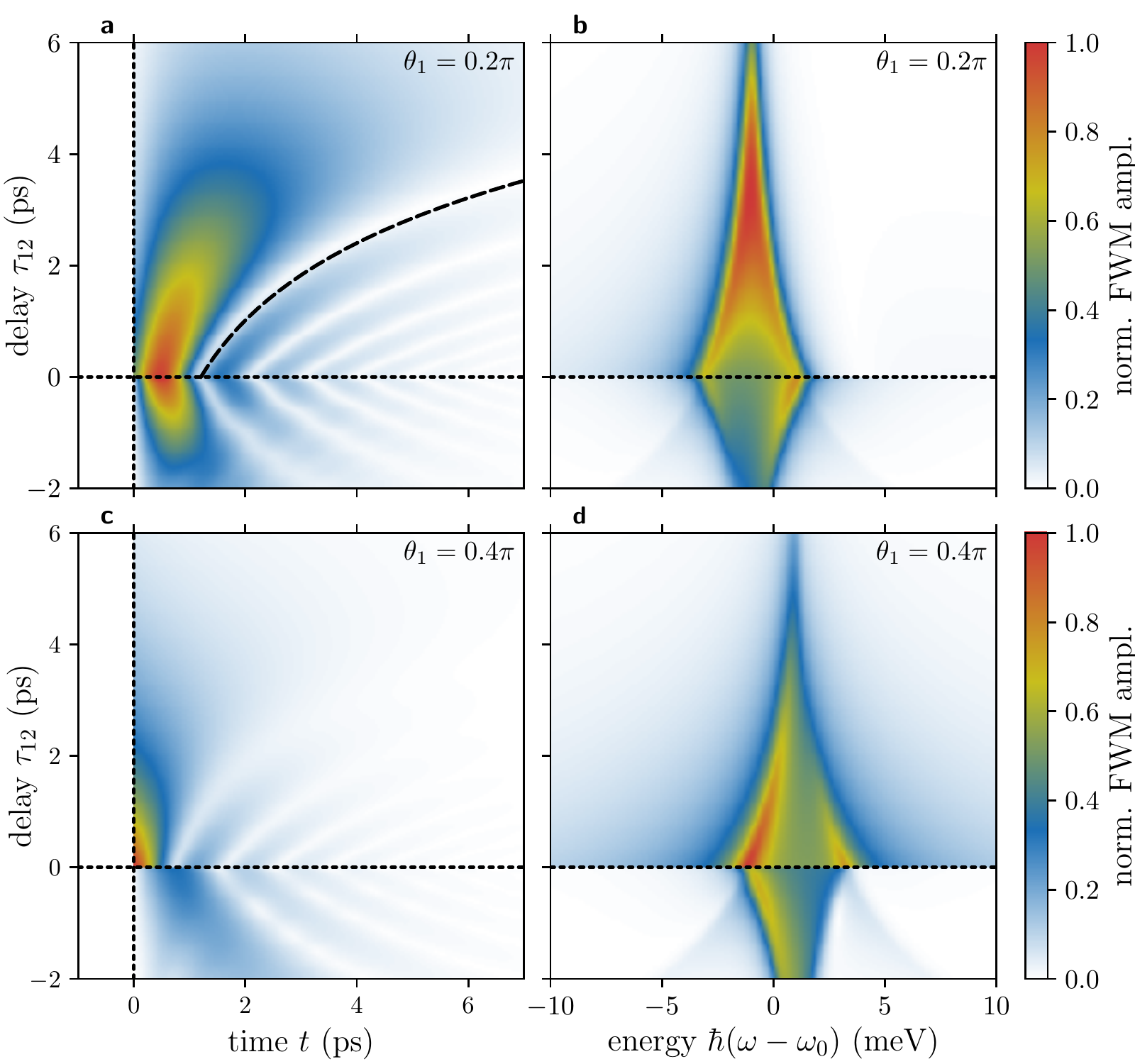}
		\caption{FWM signals for $\delta$-pulses. (a, b) $\theta_1 = 0.2\pi$ and (c, d) $\theta_1 = 0.4\pi$. FWM dynamics in (a) and (c) and the respective FWM spectra as a function of energy $\hbar(\omega-\omega_0)$ and delay $\tau_{12}$ in (b) and (d). The dashed line in (a) indicates the time dependence of the minima as predicted by Eq.~\eref{eq:minima}.}\label{fig:50}
\end{figure}

In the main text we discuss FWM signals from non-vanishing pulse durations and interpreted the signals with the help of calculations in the $\delta$-pulse limit. To get a clearer view on the impact of the pulse duration, the FWM-dynamics and spectra from Fig.~\ref{fig:10} are calculated for $\delta$-pulses via Eqs.~\eref{eq:2pfwm} and \eref{eq:neg_delay}. Regarding the case of $\theta_1 = 0.2\pi$, whose FWM dynamics and spectra are depicted in Figs.~\ref{fig:50}(a) and (b), respectively, the analytical solution matches the numerically simulated signal almost perfectly for both positive and negative delays.

For the larger pulse area $\theta_1 = 0.4\pi$ the dynamics shown in Figs.~\ref{fig:50}(c) and (d) exhibit more prominent differences with respect to Figs.~\ref{fig:10}(c) and (d). On the one hand, the oscillating signal follows the derived logarithmic behavior also for small delays because the influence of the pulse overlap vanishes for $\delta$-pulses. On the other hand, the transition from positive to negative delays is very abrupt. Nevertheless, the qualitative aspects of the spectra, i.e., width, line splitting, and narrowing  are well-reproduced. This demonstrates that overall an interpretation of the spectra via the Bloch vector retrieved in the $\delta$-pulse limit is well justified.

\begin{figure}[t]
	\includegraphics[width = 0.6 \columnwidth]{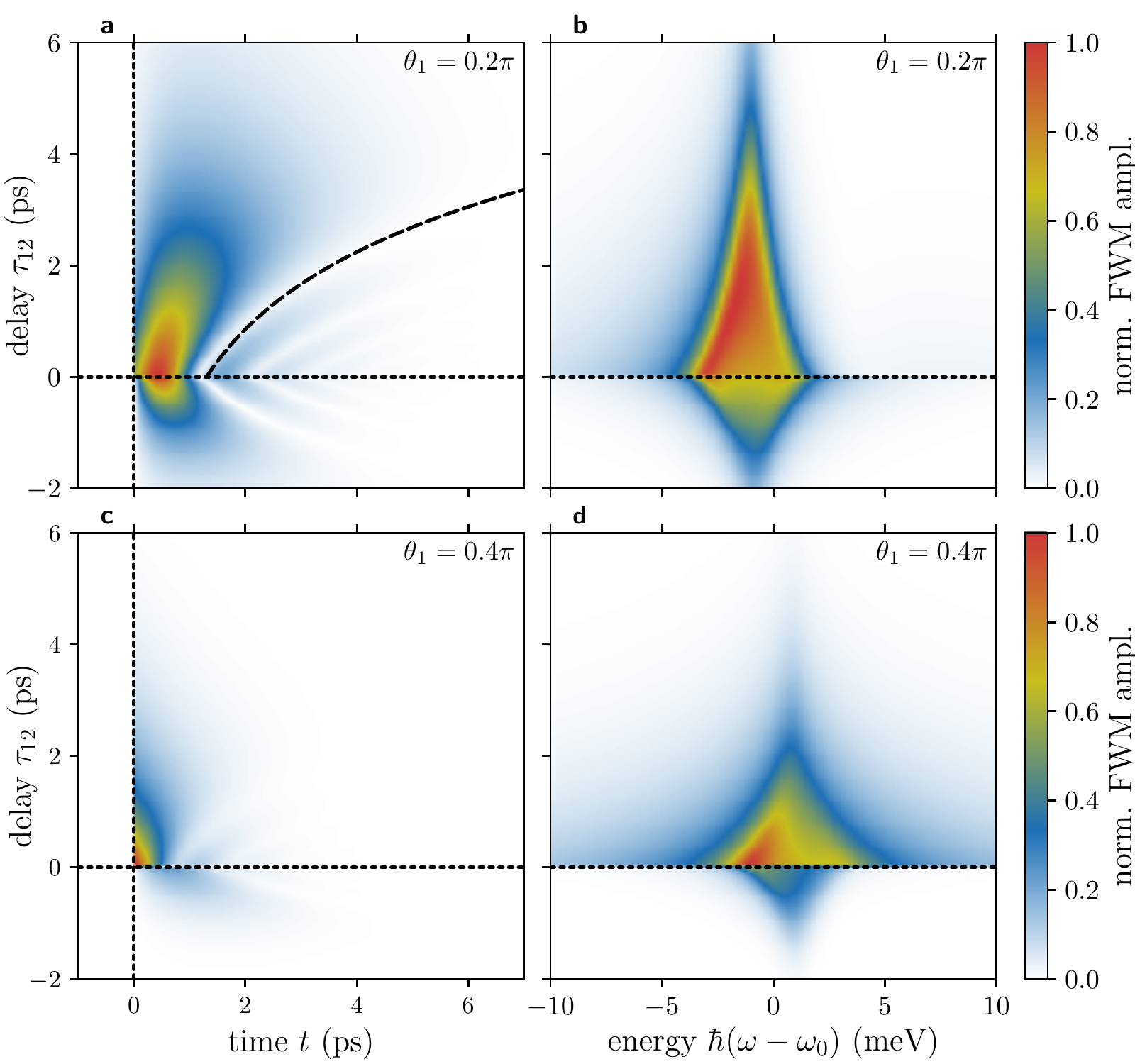}
		\caption{FWM signals in the ultrashort pulse limit ias in Fig.~\ref{fig:50} but including excitation-induced dephasing (EID). (a, b) $\theta_1 = 0.2\pi$ and (c, d) $\theta_1 = 0.4\pi$. FWM dynamics in (a) and (c) and the respective FWM spectra as a function of energy $\hbar(\omega-\omega_0)$ and delay $\tau_{12}$ in (b) and (d). The dashed line in (a) indicates the time dependence of the minima as predicted by Eq.~\eref{eq:minima}. EID is described by an imaginary part ${\rm Im} (V)=-0.5~{\rm ps}^{-1}$ and $\beta$ is adjusted to obtain the same linear dephasing as in Fig.~\ref{fig:50}.}\label{fig:51}
\end{figure}

For all the results presented up to now we have assumed a real-valued local field, i.e., a real value of $V$. As discussed in the main text, an imaginary part can be added to model the phenomenon of excitation-induced dephasing (EID). This gives rise to an additional exponential decay of the polarization depending on the pulse areas of the exciting pulses, as discussed in connection with Eq.~\eref{eq:poldyn}. In Fig.~\ref{fig:51} we show the same two-pulse FWM signals as in Fig.~\ref{fig:50}, but now with EID described by an imaginary part of ${\rm Im} (V)=-0.5~{\rm ps}^{-1}$. To obtain the same linear dephasing, we set $\beta + {\rm Im} (V)=0.5~{\rm ps}^{-1}$. In the case of weak excitation $\theta_1 = 0.2 \pi$ in Figs.~\ref{fig:51}~(a) and (b), we observe a slightly enhanced decay both as a function of the real time $t$ and the delay time $\tau_{12}$ and correspondingly a slightly broader spectrum. However, the differences are rather small. This changes strongly when increasing the pulse area to $\theta_1 = 0.4\pi$ in Figs.~\ref{fig:51} (c) and (d). Now the decay is indeed much faster and there is a considerable additional broadening of the spectrum. The enhancement of the decay is particularly strong in the case of negative delay times, which can be understood from the fact the now the stronger pulse with $\theta_2 = 0.8\pi$ arrives first and leads to a higher occupation between the pulses than in the case of positive delays. 

The results depicted in Fig.~\ref{fig:51} also show that despite the faster decay of the signals and the corresponding broadening of the spectra, the overall shape of the signals and spectra remains remarkably unaffected. In particular, the minima in the signals as a function of real time and delay time (panels (a) and (c) in  Figs.~\ref{fig:50} and \ref{fig:51}) are essentially unchanged, except for a reduction in the contrast due to the faster decay. This confirms our motivation to neglect EID in the main text, where we concentrated on the characteristic temporal and spectral features introduced by a real local field.

Let us briefly comment on the order of magnitude of the parameters chosen in our studies. Our value of $V=6$~ps$^{-1}$ leads to shifts in the spectra of the order of a few millielectronvolts. In Ref.~\cite{katsch2019tdm}, based on a microscopic theory, local fields of this order have been obtained for exciton densities of the order of $10^{12}$~cm$^{-2}$. According to Ref.~\cite{lohof2018nl} the threshold for lasing in TMDC materials, which is related to the presence of an inversion, is expected for carrier densities of the order of a few times $10^{13}$~cm$^{-2}$, indicating that densities of several $10^{12}$~cm$^{-2}$ are indeed already close to inversion. The EID parameter has been calculated to be in the range of about 1 meV for densities of the order of $10^{12}$~cm$^{-2}$~\cite{katsch2020prl}. Therefore we conclude that our parameters have a realistic order of magnitude for TMDC materials.

\section*{References}
\bibliographystyle{iopart-num}

\providecommand{\newblock}{}

\end{document}